\documentclass[12pt]{article}
\usepackage{amssymb, amsmath}
\usepackage{amsfonts}
\usepackage{color}
\newtheorem{theorem}{Theorem}[section]
\newtheorem{lemma}[theorem]{Lemma}

\newtheorem{example}[theorem]{Example}

\newtheorem{assumption}[theorem]{Assumption}

\numberwithin{equation}{section}
\def\re{\mathbb{R}}
\def\co{\mathbb{C}}

\def\ze{\mathbb{Z}}

\newenvironment{proof}[1]%
{\par\noindent{\em #1:\ }}%
{~\rule{2mm}{2mm}\par\bigskip}

\begin{document}
        \begin{center}
            {\Large\bf  Superconductivity and Low Energy Excitations\\ in an Attractive Hubbard Model\\}

            \bigskip\medskip
            {\large
                Yukimi Goto,\footnote{\small \it University of Tokyo,~Graduate School of Mathematical Sciences,
                    Komaba, Meguro-ku Tokyo 153-8914, Japan. Email:~{\tt   yukimi@ms.u-tokyo.ac.jp}}
                Tohru Koma,\footnote{\small \it Department of Physics, Gakushuin University (retired),
                    Mejiro, Toshima-ku, Tokyo 171-8588, Japan}
                Hironobu Yoshida\footnote{\small \it Nonequilibrium Quantum Statistical Mechanics RIKEN Hakubi Research Team,
                    Pioneering Research Institute (PRI), RIKEN, 2-1 Hirosawa, Wako, Saitama 351-0198, Japan.} \footnote{Present affiliations: Department of Physics, Graduate School of Science, The University of Tokyo, 7-3-1 Hongo, Tokyo 113-0033, Japan;
                    Nonequilibrium Quantum Statistical Mechanics RIKEN Hakubi Research Team,
                    Pioneering Research Institute (PRI), RIKEN, 2-1 Hirosawa, Wako, Saitama 351-0198, Japan.}
            }
        \end{center}

        \medskip
        \noindent
        {\bf Abstract:} We study an attractive Hubbard model on bipartite lattices. In the grand canonical formalism,
        we prove the existence of superconducting long-range order in the ground state on Lieb lattices
        with the chemical potential corresponding to half filling. We also study the low-energy excitations above several ground states
        for the translationally invariant Hamiltonian.
        We prove the following: (i) The pairing excitations are gapless above the ground states
        when the number of fermions deviates from that of the half filling by the order of the volume.
        (ii) A certain class of single-fermion excitations shows a non-vanishing spectral gap above the ground state with
        an even number of fermions in a strong coupling and low-density regime.
    \section{Introduction}

    As is well known, Lieb proved two theorems\;\cite{Lieb} about the ground states of the Hubbard model
    on certain bipartite lattices.\footnote{For previous results on the Hubbard model, {see~\cite{Tasaki, Mielke,LiebR}.}}
    Four years later, by relying on Lieb's results,
    Shen and Qiu\;\cite{ShenQiu} proved the existence of superconducting long-range order in the ground state
    for certain fillings of fermions in the case of the attractive interaction on the so-called Lieb lattices.
    {To the best of our knowledge, their result is the only known proof of superconductivity for the spin-$1/2$ Hubbard model.}
    However, for attractive interactions, the corresponding Lieb's theorem is restricted to
    the case of an arbitrary even number of fermions.
    Namely, the theorem is silent about the case of an odd number of fermions.
    The difference between even and odd numbers of fermions affects the statement of the theorem.

    On the other hand, Kubo and Kishi\;\cite{KK} dealt with the Hubbard model at non-zero temperatures,
    and obtained the upper bounds for the spin susceptibility in the case of the attractive interaction and
    for the charge and the on-site pairing susceptibilities in the case of the repulsive interaction.
    These upper bounds imply the absence of corresponding long-range order. In particular, the bound for the pairing
    implies the absence of superconducting long-range order. Their approach is closely related to Lieb's work
    in the sense of the spin reflection. Actually, the attractive Hubbard model exhibits the special property
    which is often called the spin reflection positivity.

    In the present paper, we deal with the Hubbard model using the grand canonical formalism \cite{Shiba},
    in order to avoid the problem of the even-oddness of the number of fermions in Lieb's theorem.
    In other words, instead of fixing the number of fermions, we introduce the chemical potential.
    We then obtain the ground-state expectation by taking the zero temperature limit.

    As a result, we prove the existence of superconducting long-range order in the ground state of
    the attractive Hubbard model on Lieb lattices. Here the system is at half filling in the sense of the expectation value of the number of fermions.
    We stress that the long-range order implies the existence of a pure state which exhibits a non-vanishing order parameter of superconductivity in the infinite-volume limit.
    {We include the proof of this implication, although we do not claim that the grand canonical treatment itself plays a crucial role in the proof.}
    The proof of long-range order is closely related to that of Shen and Qiu~\cite{ShenQiu}, however, we deal with the whole Fock space in the grand canonical formalism.
    {This extension is one of the main novelties of the present paper.
        The key technical point is to show that the sectors with an odd number of fermions do not affect the existence of superconducting long-range order.
        The comparison argument used for this purpose is also new.
        On the other hand, the proof of the long-range order bound in the even particle number sector is essentially contained in the work of Shen and Qiu~\cite{ShenQiu}, whereas they established the result for canonical ground states.}

    We also study the low-energy excitations above the ground states of the attractive Hubbard model
    at several fillings. We prove that the pairing excitations are gapless above the ground states
    when the number of fermions deviates from that of the half filling by the order of the volume.
    In addition, we prove that a certain class of single-fermion excitations has a non-vanishing spectral gap
    above the ground state with an even number of fermions in a strong coupling and low-density regime for fermions.
    The latter is a physically reasonable consequence due to the attractive pair interaction of the present model.
    Namely, the existence of an unpaired fermion always increases the whole energy.
    {The evenness of the fermion number alone does not force such a situation.}

    The present paper is organized as follows: The precise definition of the attractive Hubbard model is given in Sec.~\ref{Sec:Model}.
    The Shiba transformation \cite{Shiba}, the grand canonical formalism and the charge neutrality
    are also explained in this section. Following Kubo and Kishi \cite{KK},
    we prove the positivity of the superconducting correlation in Sec.~\ref{Sec:PosiSuper}.
    As shown in Sec.~\ref{Sec:BoundSuper}, the positivity yields some useful bounds.
    In order to use Lieb's results which are restricted to the even numbers of fermions, we consider a certain fermion Fock space
    which is restricted to the corresponding sector in Sec.~\ref{Sec:RestFock}.
    In Sec.~\ref{Sec:LROsuper}, we prove the existence of superconducting long-range order
    for the attractive Hubbard model with the chemical potential corresponding to half filling
    on the Lieb lattices (Theorem~\ref{LROsuper}). The long-range order implies the existence of a pure state which
    exhibits the non-vanishing superconducting order parameter (Theorem~\ref{sponsuper}).
    In Sec.~\ref{Sec:GapSFE}, we prove the existence of a non-vanishing spectral gap above the ground state with
    an even number of fermions in a strong coupling and low-density regime for fermions for the lattice translationally invariant
    Hamiltonian of the attractive Hubbard model on the hypercubic lattices $\ze^d$ (Theorem~\ref{thmGap}).
    The case of Lieb lattices is also treated (Theorem~\ref{thmGapLieb}).
    In Sec.~\ref{Sec:LLE}, we prove that the pairing excitations are gapless above the ground states
    when the number of fermions deviates from that of the half filling by the order of the volume
    for the lattice translationally invariant Hamiltonian (Theorems~\ref{GaplessEx} and \ref{GaplessEx2}).

    \section{Model and Properties}
    \label{Sec:Model}

    We consider an attractive Hubbard model on a bipartite lattice $\Lambda$,
    which consists of two sublattices, $\Lambda_{\rm A}$ and $\Lambda_{\rm B}$,
    i.e., $\Lambda=\Lambda_{\rm A}\cup\Lambda_{\rm B}$ and $\Lambda_{\rm A}\cap\Lambda_{\rm B}=\emptyset$.
    When $|\Lambda_{\rm A}|\ne |\Lambda_{\rm B}|$, the lattice is often called Lieb lattice.
    We assume that the size of the lattice $|\Lambda|$ is an even integer.
    The Hamiltonian is given by
    \begin{equation}
        \label{Ham}
        H^{(\Lambda)}:=\sum_{\sigma=\uparrow,\downarrow}\sum_{x,y\in \Lambda}
        t_{x,y}(c_{x,\sigma}^\dagger c_{y,\sigma}+c_{y,\sigma}^\dagger c_{x,\sigma})
        -\sum_{x\in\Lambda}U_x\Bigl(n_{x,\uparrow}-\frac{1}{2}\Bigr)\Bigl(n_{x,\downarrow}-\frac{1}{2}\Bigr),
    \end{equation}
    where $c_{x,\sigma}^\dagger$ and $c_{x,\sigma}$ are, respectively, the creation and annihilation operators
    at the site $x\in\Lambda$ for fermions with the spin degrees of freedom, $\sigma=\uparrow,\downarrow$,
    on the finite lattice $\Lambda$, and $n_{x,\sigma}:=c_{x,\sigma}^\dagger c_{x,\sigma}$
    is the number operator with the spin $\sigma$ at the site $x$;
    these fermion operators obey the anti-commutation relations,
    $$
    \{c_{x,\sigma},c_{y,\sigma'}^\dagger\}=\delta_{x,y}\delta_{\sigma,\sigma'},
    \quad \{c_{x,\sigma},c_{y,\sigma'}\}=0,\quad\mbox{and}\quad \{c_{x,\sigma}^\dagger,c_{y,\sigma'}^\dagger\}=0,
    $$
    for $x,y\in\Lambda$ and $\sigma,\sigma'=\uparrow,\downarrow$. We assume $t_{x,y}\in\re$ and $U_x>0$.
    Namely, we consider the attractive on-site interactions.
    We also assume that the hopping amplitude $t_{x,y}$ is vanishing if two sites $x,y\in\Lambda$ belong to the same sublattice,
    and that the whole lattice $\Lambda$ is connected through non-vanishing $t_{x,y}$.
    {The connectedness assumption is needed for the application of Lieb's theorem in Sec.~\ref{Sec:RestFock} and Sec.~\ref{Sec:LROsuper}.
        By contrast, the spin-reflection positivity argument in Sec.~\ref{Sec:PosiSuper} does not require this connectivity. See~\cite{Lieb,Tasaki} for further information.}

    The assumption about the hopping amplitude $t_{x,y}$ yields that the above Hamiltonian $H^{(\Lambda)}$ of (\ref{Ham}) is invariant
    under the particle-hole transformation, $c_{x,\sigma}\rightarrow \eta_xc_{x,\sigma}^\dagger$,  where
    \begin{equation}
        \label{etax}
        \eta_x:=\begin{cases}
            1 & \mbox{if \ } x\in\Lambda_{\rm A},\\
            -1 & \mbox{if \ } x\in\Lambda_{\rm B}
        \end{cases}
    \end{equation}
    for $\sigma=\uparrow,\downarrow$. The Hamiltonian is also invariant under the SU(2) spin rotations.

    \subsection{Shiba Transformation and Interaction Hamiltonian}
    \label{Sec:Shiba}

    We introduce a useful transformation,
    \begin{equation}
        \label{Shibatrans}
        c_{x,\downarrow}\rightarrow [U_{\rm S}^{(\Lambda)}]^\dagger c_{x,\downarrow} U_{\rm S}^{(\Lambda)}=
        \eta_x c_{x,\downarrow}^\dagger\quad \mbox{and}
        \quad c_{x,\uparrow}\rightarrow [U_{\rm S}^{(\Lambda)}]^\dagger c_{x,\uparrow}U_{\rm S}^{(\Lambda)}=c_{x,\uparrow},
    \end{equation}
    for all sites, $x\in\Lambda$, which is often called Shiba transformation \cite{Shiba}.
    Here, $\eta_x$ is given by (\ref{etax}), and we have written $U_{\rm S}^{(\Lambda)}$ for the corresponding unitary transformation.
    In comparison with the particle-hole transformation, the operators for the up spin are not altered.
    Clearly, the transformation $U_{\rm S}^{(\Lambda)}$ does not change the hopping term $H_{\rm hop}^{(\Lambda)}$
    of the Hamiltonian $H^{(\Lambda)}$ of (\ref{Ham}).

    In order to see that the transformation $U_{\rm S}^{(\Lambda)}$ converts the attractive interactions to the repulsive ones,
    we write
    \begin{equation}
        \label{Hint}
        H_{\rm int}^{(\Lambda)}=-\sum_{x\in\Lambda}U_x\Bigl(n_{x,\uparrow}-\frac{1}{2}\Bigr)\Bigl(n_{x,\downarrow}-\frac{1}{2}\Bigr)
    \end{equation}
    for the attractive interaction Hamiltonian of $H^{(\Lambda)}$. Since one has
    $[U_{\rm S}^{(\Lambda)}]^\dagger n_{x,\downarrow}U_{\rm S}^{(\Lambda)}=1-n_{x,\downarrow}$ for all $x\in\Lambda$,
    the transformation yields the repulsive interactions,
    \begin{equation}
        \check{H}_{\rm int}^{(\Lambda)}:=[U_{\rm S}^{(\Lambda)}]^\dagger H_{\rm int}^{(\Lambda)} U_{\rm S}^{(\Lambda)}=
        \sum_{x\in\Lambda}U_x\Bigl(n_{x,\uparrow}-\frac{1}{2}\Bigr)\Bigl(n_{x,\downarrow}-\frac{1}{2}\Bigr).
    \end{equation}
    This is still invariant under the SU(2) spin rotations. We write
    \begin{equation}
        \check{H}^{(\Lambda)}:=[U_{\rm S}^{(\Lambda)}]^\dagger H^{(\Lambda)} U_{\rm S}^{(\Lambda)}.
        \label{tildeHam}
    \end{equation}

    \subsection{Grand Canonical Formalism and Charge Neutrality}
    \label{Sec:GCF}

    First, we consider the grand canonical ensemble for the present model. The thermal expectation value is given by
    \begin{equation}
        \langle\cdots\rangle_{\beta,\Lambda}:=\frac{1}{Z_{\beta,\Lambda}}{\rm Tr}\;(\cdots)\exp[-\beta H^{(\Lambda)}],
    \end{equation}
    where {the trace is taken over the fermion Fock space,} $\beta$ is the inverse temperature, and $Z_{\beta,\Lambda}:={\rm Tr} \exp[-\beta H^{(\Lambda)}]$.
    Here, we have chosen the chemical potential to be zero so that the half filling is realized for the particle density.
    To see this, we write
    \begin{equation}
        n^{(\Lambda)}:=\sum_{x\in\Lambda} n_x,
    \end{equation}
    where $n_x:=n_{x,\uparrow}+n_{x,\downarrow}$. By using the particle-hole symmetry, one has
    \begin{equation}
        \langle n^{(\Lambda)}\rangle_{\beta,\Lambda}=2|\Lambda|-\langle n^{(\Lambda)}\rangle_{\beta,\Lambda}.
    \end{equation}
    This yields
    \begin{equation}
        \frac{1}{|\Lambda|}\langle n^{(\Lambda)}\rangle_{\beta,\Lambda}=1.
    \end{equation}

    In order to neutralize the total charge,
    a fictitious background positive charge is often added to the system with a fixed chemical potential.
    However, this prescription does not always work well. In fact, it was proved \cite{YK} that
    an attractive SU($N$) Hubbard model with flavor $N\ge 3$ exhibits the particle-hole symmetry breaking,
    and consequently there appear two ground states which show different filling from the expected one.
    Clearly, there arises a question whether or not there exists an infinite-volume pure ground state which exhibits
    the half filling in the sense of the expectation value of the charge density.

    \section{Positivity and Bounds for the Correlation}
    {In this section, we derive a useful bound on the superconducting correlation function. For this purpose, we first establish positivity of the correlation in the next subsection.}
    \subsection{Positivity of Superconducting Correlation}
    \label{Sec:PosiSuper}

    By using the spin reflection positivity \cite{KK}, we show below the positivity (\ref{superpositive})
    of the correlation of superconductivity.
    {The result follows directly from the arguments in~\cite{KK}, but we provide a proof for the readers' convenience.}
    To begin with, note that
    \begin{equation}
        (n_{x,\uparrow}-n_{x,\downarrow})^2= (n_{x,\uparrow}-n_{x,\downarrow})(n_{x,\uparrow}-n_{x,\downarrow})
        =n_{x,\uparrow}+n_{x,\downarrow}-2n_{x,\uparrow}n_{x,\downarrow}
    \end{equation}
    and
    \begin{eqnarray*}
        \left(n_{x,\uparrow}-\frac{1}{2}\right)\left(n_{x,\downarrow}-\frac{1}{2}\right)
        &=& n_{x,\uparrow}n_{x,\downarrow}-\frac{1}{2}n_{x,\uparrow}-\frac{1}{2}n_{x,\downarrow}+\frac{1}{4}\\
        &=&\frac{1}{2}\left(2n_{x,\uparrow}n_{x,\downarrow}-n_{x,\uparrow}-n_{x,\downarrow}\right)+\frac{1}{4}.
    \end{eqnarray*}
    These yield
    \begin{equation}
        \left(n_{x,\uparrow}-\frac{1}{2}\right)\left(n_{x,\downarrow}-\frac{1}{2}\right)
        =-\frac{1}{2}\left(n_{x,\uparrow}-n_{x,\downarrow}\right)^2+\frac{1}{4}.
    \end{equation}
    By using this identity, the interaction Hamiltonian $H_{\rm int}^{(\Lambda)}$ of (\ref{Hint}) can be written
    \begin{equation}
        \label{Hint2}
        H_{\rm int}^{(\Lambda)}=\sum_{x\in\Lambda}\frac{U_x}{2}\left(n_{x,\uparrow}-n_{x,\downarrow}\right)^2
        -\frac{1}{4}\sum_{x\in\Lambda}U_x.
    \end{equation}
    Therefore, by using Lie product formula, one has \cite{KK}
    \begin{eqnarray}
        & &\exp[-\beta H^{(\Lambda)}-\beta\mathcal{C}_0]\nonumber\\
        &=&\lim_{M\nearrow\infty}\left\{\exp\left[-\frac{\beta}{M} H_{{\rm hop},\uparrow}^{(\Lambda)}\right]
        \exp\left[-\frac{\beta}{M} H_{{\rm hop},\downarrow}^{(\Lambda)}\right]
        \exp\left[-\frac{\beta}{2M}\sum_{x\in\Lambda}U_x(n_{x,\uparrow}-n_{x,\downarrow})^2\right]\right\}^M,\nonumber\\
        \label{Lieprod}
    \end{eqnarray}
    where we have written
    \begin{equation}
        H_{{\rm hop},\sigma}^{(\Lambda)}:=\sum_{x,y\in \Lambda}
        t_{x,y}(c_{x,\sigma}^\dagger c_{y,\sigma}+c_{y,\sigma}^\dagger c_{x,\sigma})
        \quad \mbox{for \ } \sigma=\uparrow,\downarrow,
    \end{equation}
    and the constant is given by
    \begin{equation}
        \mathcal{C}_0:=\frac{1}{4}\sum_{x\in\Lambda}U_x.
    \end{equation}
    By using the operator identity,
    \begin{equation}
        \label{opeID}
        e^{-D^2}=\frac{1}{(4\pi)^{1/2}}\int_{\mathbb{R}} dk \exp\left[-\frac{1}{4}k^2+ikD\right],
    \end{equation}
    for a hermitian $D$, one has
    \begin{equation}
        \exp\left[-\frac{\beta U_x}{2M}(n_{x,\uparrow}-n_{x,\downarrow})^2\right]
        =\frac{1}{(4\pi)^{1/2}}\int_{\mathbb{R}}  dk \exp\left[-k^2/4\right]
        \exp\left[ik\sqrt{\frac{\beta U_x}{2M}}(n_{x,\uparrow}-n_{x,\downarrow})\right],
    \end{equation}
    where we have used the assumption $U_x>0$.
    {All integrations with respect to the real variables in this section are over the entire real line, and we omit the limits of integration below.}
    By applying this identity to the right-hand side of (\ref{Lieprod}), we have
    \begin{eqnarray}
        & &\exp[-\beta H^{(\Lambda)}-\beta\mathcal{C}_0]\nonumber\\
        &=&\lim_{M\nearrow\infty}\int \cdots \int \Bigl[\prod_{x\in\Lambda}\prod_{\ell=1}^M d\mu(k_{x,\ell})\Bigr]
        \alpha_{1,\uparrow}^{(M)}\alpha_{2,\uparrow}^{(M)}\cdots \alpha_{M,\uparrow}^{(M)}\cdot
        \overline{\alpha_{1,\downarrow}^{(M)}}\; \overline{\alpha_{2,\downarrow}^{(M)}}\cdots \overline{\alpha_{M,\downarrow}^{(M)}},
        \label{expbetaHalpha}
    \end{eqnarray}
    where we have written
    \begin{equation}
        \alpha_{\ell,\sigma}^{(M)}:=\exp\left[-\frac{\beta}{M}H_{{\rm hop},\sigma}^{(\Lambda)}\right]
        \prod_{x\in\Lambda}\exp\left[ik_{x,\ell}\sqrt{\frac{\beta U_x}{2M}}n_{x,\sigma}\right]
    \end{equation}
    for $\sigma=\uparrow,\downarrow$, and $\ell=1,2,\ldots,M$; the measure is given by
    \begin{equation}
        d\mu(k):=\frac{dk}{(4\pi)^{1/2}}\exp[-k^2/4],
    \end{equation}
    and $\overline{\cdots}$ denotes the complex conjugate.

    In order to study the superconducting long-range order,
    let us consider the observable $c_{x,\uparrow}^\dagger c_{x,\downarrow}^\dagger c_{y,\downarrow}c_{y,\uparrow}$
    for $x,y\in\Lambda$. Clearly, one has
    \begin{equation}
        c_{x,\uparrow}^\dagger c_{x,\downarrow}^\dagger c_{y,\downarrow}c_{y,\uparrow}
        =c_{x,\uparrow}^\dagger c_{y,\uparrow}\cdot c_{x,\downarrow}^\dagger c_{y,\downarrow}
    \end{equation}
    for any $x,y\in\Lambda$.
    {Now we use the fact that the up-spin and down-spin Fock spaces are isomorphic, and their traces obey ${\rm Tr}_{\downarrow}(A_{\downarrow})={\rm Tr}_{\uparrow}(A_{\uparrow})$ whenever $A_{\downarrow}$ is obtained from $A_{\uparrow}$ by interchanging the up and down spins.
        Then we deduce from the expression $\exp[-\beta H^{(\Lambda)}]$ of (\ref{expbetaHalpha}) that
        \[
        {\rm Tr}\left[c_{x,\uparrow}^\dagger c_{y,\uparrow}\cdot c_{x,\downarrow}^\dagger c_{y,\downarrow}\exp(-\beta H^{(\Lambda)})\right]
        =e^{\beta \mathcal{C}_0}
        \lim_{M\nearrow\infty}\int d\tilde \mu\,\mathrm{Tr} \left[ c_{x,\uparrow}^\dagger c_{y,\uparrow} \cdot c_{x,\downarrow}^\dagger c_{y,\downarrow}\prod_{i=1}^{M} \alpha^{(M)}_{i,\uparrow} \overline{\alpha^{(M)}_{i,\downarrow}}\right],
        \]
        where $d\tilde \mu :=\prod_{x\in\Lambda}\prod_{\ell=1}^M d\mu(k_{x,\ell})$.
        Since the matrix elements of fermion operators $\{c_{x,\sigma}\}$ can be chosen to be real and all the $\alpha^{(M)}_{i,\sigma}$ consist of even operators, we have
        \begin{align*}
            \mathrm{Tr} \left[ c_{x,\uparrow}^\dagger c_{y,\uparrow} \cdot c_{x,\downarrow}^\dagger c_{y,\downarrow}\prod_{i=1}^{M} \alpha^{(M)}_{i,\uparrow} \overline{\alpha^{(M)}_{i,\downarrow}}  \right]
            &=
            \mathrm{Tr}_{\uparrow} \left[ c_{x,\uparrow}^\dagger c_{y,\uparrow} \prod_{i=1}^{M} \alpha^{(M)}_{i,\uparrow}\right]
            \cdot  \mathrm{Tr}_{\downarrow} \left[ c_{x,\downarrow}^\dagger c_{y,\downarrow} \prod_{i=1}^{M} \overline{\alpha^{(M)}_{i,\downarrow}}\right]\\
            &=
            \left| \mathrm{Tr}_{\uparrow} \left[ c_{x,\uparrow}^\dagger c_{y,\uparrow} \prod_{i=1}^{M} \alpha^{(M)}_{i,\uparrow}\right]
            \right|^2.
        \end{align*}
        This proves
    }
    \begin{equation}
        {\rm Tr}\left[c_{x,\uparrow}^\dagger c_{y,\uparrow}\cdot c_{x,\downarrow}^\dagger c_{y,\downarrow}\exp(-\beta H^{(\Lambda)})\right]\ge 0,
    \end{equation}
    which means that
    \begin{equation}
        \label{superpositive}
        \langle c_{x,\uparrow}^\dagger c_{x,\downarrow}^\dagger c_{y,\downarrow}c_{y,\uparrow}\rangle_{\beta,\Lambda}\ge 0
        \quad \mbox{for any } x,y\in\Lambda.
    \end{equation}

    \subsection{Bounds for Superconducting Correlation}
    \label{Sec:BoundSuper}

    By using Shiba transformation $U_{\rm S}^{(\Lambda)}$, we have
    \begin{equation}
        \label{superShiba}
        [U_{\rm S}^{(\Lambda)}]^\dagger c_{x,\uparrow}^\dagger c_{x,\downarrow}^\dagger c_{y,\downarrow}c_{y,\uparrow}U_{\rm S}^{(\Lambda)}
        =\eta_x\eta_y c_{x,\uparrow}^\dagger c_{x,\downarrow} c_{y,\downarrow}^\dagger c_{y,\uparrow}
        =\eta_x\eta_y S_x^+S_y^-,
    \end{equation}
    where we have written $S_x^+:=c_{x,\uparrow}^\dagger c_{x,\downarrow}$ and $S_y^-:=c_{y,\downarrow}^\dagger c_{y,\uparrow}$.
    We also write
    \begin{equation}
        S_x^{(i)}:=\frac{1}{2}(c_{x,\uparrow}^\dagger,c_{x,\downarrow}^\dagger)\tau^{(i)}
        \begin{pmatrix}
            c_{x,\uparrow} \\ c_{x,\downarrow}
        \end{pmatrix} \ \ \mbox{for} \ \ i=1,2,3,
    \end{equation}
    where
    \begin{equation}
        \tau^{(1)}=
        \begin{pmatrix}
            0 & 1 \\
            1 & 0
        \end{pmatrix},\ \
        \tau^{(2)}=
        \begin{pmatrix}
            0 & -i \\
            i & 0
        \end{pmatrix}
        \ \ \mbox{and} \ \
        \tau^{(3)}=
        \begin{pmatrix}
            1 & 0 \\
            0 & -1
        \end{pmatrix}.
    \end{equation}
    Immediately, one has $S_x^\pm=S_x^{(1)}\pm iS_x^{(2)}$ and $S_x^{(3)}=(n_{x,\uparrow}-n_{x,\downarrow})/2$.
    Therefore, by using the commutation relation $[S_x^{(1)},S_y^{(2)}]=i\delta_{x,y}S_x^{(3)}$, we have
    \begin{equation}
        S_x^+S_y^-+S_y^+S_x^-=2[S_x^{(1)}S_y^{(1)}+S_x^{(2)}S_y^{(2)}+\delta_{x,y}S_x^{(3)}].
    \end{equation}
    Further, from (\ref{superShiba}), the expectation value of the superconducting correlation is written
    \begin{eqnarray}
        & &\langle c_{x,\uparrow}^\dagger c_{x,\downarrow}^\dagger c_{y,\downarrow}c_{y,\uparrow}\rangle_{\beta,\Lambda}
        +\langle c_{y,\uparrow}^\dagger c_{y,\downarrow}^\dagger c_{x,\downarrow}c_{x,\uparrow}\rangle_{\beta,\Lambda}\nonumber\\
        &=&2\eta_x\eta_y[\langle\!\langle S_x^{(1)}S_y^{(1)}\rangle\!\rangle_{\beta,\Lambda}
        +\langle\!\langle S_x^{(2)}S_y^{(2)}\rangle\!\rangle_{\beta,\Lambda}
        +\delta_{x,y}\langle\!\langle S_x^{(3)}\rangle\!\rangle_{\beta,\Lambda}],
        \label{superspinexpc}
    \end{eqnarray}
    where we have written
    \begin{equation}
        \langle\!\langle \cdots \rangle\!\rangle_{\beta,\Lambda}:=\frac{1}{\check{Z}_{\beta,\Lambda}}{\rm Tr}\;(\cdots)
        e^{-\beta \check{H}^{(\Lambda)}}
    \end{equation}
    for the Hamiltonian $\check{H}^{(\Lambda)}$ of (\ref{tildeHam}) with the partition function
    $\check{Z}_{\beta,\Lambda}:={\rm Tr}\exp[-\beta \check{H}^{(\Lambda)}]$.
    Since the Hamiltonian $\check{H}^{(\Lambda)}$ is invariant under the SU(2) spin rotation, one has
    \begin{equation}
        \langle c_{x,\uparrow}^\dagger c_{x,\downarrow}^\dagger c_{y,\downarrow}c_{y,\uparrow}\rangle_{\beta,\Lambda}
        +\langle c_{y,\uparrow}^\dagger c_{y,\downarrow}^\dagger c_{x,\downarrow}c_{x,\uparrow}\rangle_{\beta,\Lambda}
        =4\eta_x\eta_y \langle\!\langle S_x^{(3)}S_y^{(3)}\rangle\!\rangle_{\beta,\Lambda}\ge
        4\left| \langle\!\langle S_x^{(3)}S_y^{(3)}\rangle\!\rangle_{\beta,\Lambda} \right|.
    \end{equation}
    The inequality follows from the positivity (\ref{superpositive}) of the superconducting correlation \cite{ShenQiu,SQT}.
    Moreover, by using Shiba transformation (\ref{Shibatrans}) again, we have
    \begin{equation}
        \label{superchargeCorr}
        \begin{split}
            \langle c_{x,\uparrow}^\dagger c_{x,\downarrow}^\dagger c_{y,\downarrow}c_{y,\uparrow}\rangle_{\beta,\Lambda}
            +\langle c_{y,\uparrow}^\dagger c_{y,\downarrow}^\dagger c_{x,\downarrow}c_{x,\uparrow}\rangle_{\beta,\Lambda}
            &=\eta_x\eta_y \langle (n_x-1)(n_y-1)\rangle_{\beta,\Lambda}\\
            &= \left|\langle (n_x-1)(n_y-1)\rangle_{\beta,\Lambda}\right|\\
            &\ge
            \langle (n_x-1)(n_y-1)\rangle_{\beta,\Lambda},
        \end{split}
    \end{equation}
    where $n_x=n_{x,\uparrow}+n_{x,\downarrow}$. By summing over all the sites, $x,y\in\Lambda$, we can obtain
    the desired result,
    \begin{equation}
        \label{boudsuper}
        \langle[O_{\rm super}^{(\Lambda)}]^\dagger O_{\rm super}^{(\Lambda)}\rangle_{\beta,\Lambda}
        =\frac{1}{2}\langle [O_{\rm CDW}^{(\Lambda)}]^2\rangle_{\beta,\Lambda}
        \ge \frac{1}{2}\langle [\delta\rho^{(\Lambda)}]^2\rangle_{\beta,\Lambda},
    \end{equation}
    where $O_{\rm super}^{(\Lambda)}$, $O_{\rm CDW}^{(\Lambda)}$ and $\delta\rho^{(\Lambda)}$ are
    the order parameters for the superconductivity, the charge-density wave and the deviation of the charge density from the half filling,
    respectively; the precise expressions are given by
    \begin{equation}
        \label{Os}
        O_{\rm super}^{(\Lambda)}:=\frac{1}{|\Lambda|}\sum_{x\in\Lambda}c_{x,\downarrow}c_{x,\uparrow},\ \
        O_{\rm CDW}^{(\Lambda)}:=\frac{1}{|\Lambda|}\sum_{x\in\Lambda}\eta_x(n_{x}-1), \ \ \mbox{and} \ \
        \delta\rho^{(\Lambda)}:=\frac{1}{|\Lambda|}\sum_{x\in\Lambda}(n_x-1).
    \end{equation}

    \section{Restriction of Fermion Fock Space}
    \label{Sec:RestFock}

    In Lieb's theorems \cite{Lieb}, the statement about the attractive Hubbard model is restricted to
    an even number of fermions. In order to use the result, we want to restrict the fermion Fock space
    to the corresponding sector.

    To begin with, we write $N_\sigma$ for the number of fermions with spin $\sigma\in\{\uparrow,\downarrow\}$.
    By using the even-oddness of $N_\sigma$,
    we decompose the trace, ${\rm Tr}\;(\cdots)$, into four parts as follows:
    \begin{equation}
        {\rm Tr}\;(\cdots)={\rm Tr}^{\rm e,e}(\cdots)+{\rm Tr}^{\rm o,o}(\cdots)+{\rm Tr}^{\rm e,o}(\cdots)+{\rm Tr}^{\rm o,e}(\cdots).
    \end{equation}
    The first term ${\rm Tr}^{\rm e,e}(\cdots)$ on the right-hand side denotes the restriction of the trace to
    the sector of the fermion Fock space with $N_\uparrow$=even and $N_\downarrow$=even. Similarly,
    ${\rm Tr}^{\rm o,o}(\cdots)$, ${\rm Tr}^{\rm e,o}(\cdots)$ and ${\rm Tr}^{\rm o,e}(\cdots)$ denote
    three parts, $\{N_\uparrow,N_\downarrow\}=\{\{{\rm odd},{\rm odd}\},\{{\rm even},{\rm odd}\},\{{\rm odd},{\rm even}\}\}$,
    respectively.
    For $N=N_\uparrow+N_\downarrow= {\rm even}$, we define the expectation value by
    \begin{equation}
        \langle\cdots\rangle_{\beta,\Lambda}^{\rm even}:=\frac{1}{Z_{\beta,\Lambda}^{\rm even}}
        \left[{\rm Tr}^{\rm e,e}(\cdots)e^{-\beta H^{(\Lambda)}}+{\rm Tr}^{\rm o,o}(\cdots)e^{-\beta H^{(\Lambda)}}\right],
    \end{equation}
    where
    \begin{equation}
        Z_{\beta,\Lambda}^{\rm even}:={\rm Tr}^{\rm e,e}e^{-\beta H^{(\Lambda)}}+{\rm Tr}^{\rm o,o}e^{-\beta H^{(\Lambda)}}.
    \end{equation}
    Then, we have

    \begin{lemma}
        The following bound is valid:
        \begin{equation}
            \label{Osuperbound}
            \langle [O_{\rm super}^{(\Lambda)}]^\dagger O_{\rm super}^{(\Lambda)}\rangle_{\beta,\Lambda}
            \ge \frac{1}{2}\langle[O_{\rm super}^{(\Lambda)}]^\dagger O_{\rm super}^{(\Lambda)}\rangle_{\beta,\Lambda}^{\rm even}
        \end{equation}
        for the order parameter $O_{\rm super}^{(\Lambda)}$ of superconductivity in (\ref{Os}).
    \end{lemma}

    \begin{proof}{Proof}
        In order to prove the above bound (\ref{Osuperbound}), we recall the expression (\ref{expbetaHalpha}),
        \begin{equation}
            \exp[-\beta H^{(\Lambda)}-\beta\mathcal{C}_0]=\lim_{M\nearrow \infty}\int d\tilde{\mu}\;
            \tilde{\alpha}_\uparrow^{(M)}\overline{\tilde{\alpha}_\downarrow^{(M)}},
        \end{equation}
        where we have written
        \begin{equation}
            \tilde{\alpha}_\sigma^{(M)}:=\alpha_{1,\sigma}^{(M)}\alpha_{2,\sigma}^{(M)}\cdots \alpha_{M,\sigma}^{(M)}
        \end{equation}
        for $\sigma\in\{\uparrow,\downarrow \}$, and
        \begin{equation}
            d\tilde{\mu}:=\prod_{x\in\Lambda}\prod_{\ell=1}^M d\mu(k_{x,\ell}).
        \end{equation}
        We have
        \begin{eqnarray}
            {\rm Tr}^{\rm e,o}(e^{-\beta H^{(\Lambda)}})&=&e^{\beta\mathcal{C}_0}\lim_{M\nearrow \infty}\int d\tilde{\mu}\;
            {\rm Tr}^{\rm e,o}(\tilde{\alpha}_\uparrow^{(M)}\overline{\tilde{\alpha}_\downarrow^{(M)}})\nonumber\\
            &=&e^{\beta\mathcal{C}_0}\lim_{M\nearrow \infty}\int d\tilde{\mu}\;
            {\rm Tr}_\uparrow^{\rm e}(\tilde{\alpha}_\uparrow^{(M)})\cdot\overline{{\rm Tr}_\downarrow^{\rm o}(\tilde{\alpha}_\downarrow^{(M)})},
        \end{eqnarray}
        where ${\rm Tr}_\sigma^{\rm p}(\cdots)$ is the restriction of the trace to the sector with
        the number $N_\sigma={\rm p}$ of fermions with spin $\sigma\in\{\uparrow,\downarrow\}$
        and ${\rm p}\in\{{\rm even},{\rm odd}\}$, and {we have used the fact that the matrix elements of the fermion creation and annihilation operators can be chosen real.}
        Note that
        \begin{eqnarray}
            \left|{\rm Tr}_\uparrow^{\rm e}(\tilde{\alpha}_\uparrow^{(M)})
            \cdot\overline{{\rm Tr}_\downarrow^{\rm o}(\tilde{\alpha}_\downarrow^{(M)})}\right|
            &\le& \frac{1}{2}\left[
            \left|{\rm Tr}_\uparrow^{\rm e}(\tilde{\alpha}_\uparrow^{(M)})\right|^2
            +\left|{\rm Tr}_\downarrow^{\rm o}(\tilde{\alpha}_\downarrow^{(M)})\right|^2\right]\nonumber\\
            &=&\frac{1}{2}\left[{\rm Tr}_\uparrow^{\rm e}(\tilde{\alpha}_\uparrow^{(M)})
            \cdot\overline{{\rm Tr}_\downarrow^{\rm e}(\tilde{\alpha}_\downarrow^{(M)})}
            +{\rm Tr}_\uparrow^{\rm o}(\tilde{\alpha}_\uparrow^{(M)})\cdot
            \overline{{\rm Tr}_\downarrow^{\rm o}(\tilde{\alpha}_\downarrow^{(M)})}\right],\nonumber\\
        \end{eqnarray}
        where we have used ${\rm Tr}_\uparrow^{\rm p}(\tilde{\alpha}_\uparrow^{(M)})
        ={\rm Tr}_\downarrow^{\rm p}(\tilde{\alpha}_\downarrow^{(M)})$ for ${\rm p}\in\{{\rm even, odd}\}$.
        By substituting this into the above right-hand side, we obtain
        \begin{equation}
            0\le {\rm Tr}^{\rm e,o}(e^{-\beta H^{(\Lambda)}})\le \frac{1}{2}\left[{\rm Tr}^{\rm e,e}(e^{-\beta H^{(\Lambda)}})
            +{\rm Tr}^{\rm o,o}(e^{-\beta H^{(\Lambda)}})\right].
        \end{equation}
        Similarly,
        \begin{equation}
            0\le {\rm Tr}^{\rm o,e}(e^{-\beta H^{(\Lambda)}})\le \frac{1}{2}\left[{\rm Tr}^{\rm e,e}(e^{-\beta H^{(\Lambda)}})
            +{\rm Tr}^{\rm o,o}(e^{-\beta H^{(\Lambda)}})\right].
        \end{equation}
        These yield
        \begin{equation}
            \label{Zboundeven}
            {\rm Tr}(e^{-\beta H^{(\Lambda)}})\le 2\left[{\rm Tr}^{\rm e,e}(e^{-\beta H^{(\Lambda)}})
            +{\rm Tr}^{\rm o,o}(e^{-\beta H^{(\Lambda)}})\right].
        \end{equation}

        On the other hand, we can show
        \begin{equation}
            {\rm Tr}^{\rm e,o}([O_{\rm super}^{(\Lambda)}]^\dagger O_{\rm super}^{(\Lambda)}e^{-\beta H^{(\Lambda)}})\ge 0
            \quad \mbox{and}\quad {\rm Tr}^{\rm o,e}([O_{\rm super}^{(\Lambda)}]^\dagger O_{\rm super}^{(\Lambda)}e^{-\beta H^{(\Lambda)}})\ge 0.
        \end{equation}
        In order to prove these inequalities, we write
        \begin{equation}
            \{\Phi_i^{(\Lambda)}(N_\uparrow,N_\downarrow)\}_i
        \end{equation}
        for the orthonormal basis of the eigenstates of the Hamiltonian $H^{(\Lambda)}$.
        Here, we have used the fact that the number $N_\sigma$ of the fermions with spin $\sigma\in\{\uparrow,\downarrow\}$ is conserved.
        Therefore, one has
        \begin{eqnarray}
            & &{\rm Tr}^{\rm e,o}([O_{\rm super}^{(\Lambda)}]^\dagger O_{\rm super}^{(\Lambda)}e^{-\beta H^{(\Lambda)}})\nonumber\\
            &=&\sum_{i:N_\uparrow={\rm even},N_\downarrow={\rm odd}}\langle \Phi_i^{(\Lambda)}(N_\uparrow,N_\downarrow),
            [O_{\rm super}^{(\Lambda)}]^\dagger O_{\rm super}^{(\Lambda)}\Phi_i^{(\Lambda)}(N_\uparrow,N_\downarrow)\rangle
            e^{-\beta E_i^{(\Lambda)}},
        \end{eqnarray}
        where $E_i^{(\Lambda)}$ is the energy eigenvalue of the Hamiltonian $H^{(\Lambda)}$. Since one has
        $$\langle \Phi_i^{(\Lambda)}(N_\uparrow,N_\downarrow),
        [O_{\rm super}^{(\Lambda)}]^\dagger O_{\rm super}^{(\Lambda)}\Phi_i^{(\Lambda)}(N_\uparrow,N_\downarrow)\rangle\ge 0,$$
        the first inequality is derived. Similarly, the second one is obtained.
        By combining these inequalities with the above bound (\ref{Zboundeven}), we can obtain the desired bound (\ref{Osuperbound}).
    \end{proof}

    For the ground states, the expectation values are given by
    \begin{equation}
        \omega_{0,\Lambda}(\cdots):=\lim_{\beta\nearrow\infty}\langle\cdots\rangle_{\beta,\Lambda}
    \end{equation}
    and
    \begin{equation}
        \omega_{0,\Lambda}^{\rm even}(\cdots):=\lim_{\beta\nearrow\infty}\langle\cdots\rangle_{\beta,\Lambda}^{\rm even}.
    \end{equation}
    Therefore, we have
    \begin{equation}
        \label{superLRObound}
        \omega_{0,\Lambda}([O_{\rm super}^{(\Lambda)}]^\dagger O_{\rm super}^{(\Lambda)})\ge \frac{1}{2}
        \omega_{0,\Lambda}^{\rm even}([O_{\rm super}^{(\Lambda)}]^\dagger O_{\rm super}^{(\Lambda)})
    \end{equation}
    for the ground states. Thus, in order to prove the existence of superconducting long-range order,
    it suffices to show it in the ground state restricted to the sector with even fermion number $N$.

    \subsection*{Ground States for an Even Number of Fermions}
    \label{sec:GSevenrepl}

    We recall Lieb's theorem \cite{Lieb} for the attractive Hubbard model. Let $N$ be an even fermion number.
    Then the ground state of the Hamiltonian $H^{(\Lambda)}$ is unique, and has the total spin $S_{\rm tot}=0$.
    Therefore, $N_\uparrow=N_\downarrow$ and $N=2N_\uparrow$. By the Shiba transformation, the numbers, $N_\uparrow$ and $N_\downarrow$,
    of the up and down spin fermions are mapped to $N_\uparrow$ and $|\Lambda|-N_\uparrow$ fermions, respectively,
    in the corresponding repulsive Hubbard model. Therefore, the total number of fermions in the repulsive Hubbard model is given by
    $|\Lambda|$.

    On the other hand, Lieb's theorem about the repulsive Hubbard model states that, when the number of fermions is given by $|\Lambda|$,
    any ground state of the Hamiltonian has the total spin $S_{\rm tot}=\left||\Lambda_{\rm A}|-|\Lambda_{\rm B}|\right|/2$.
    Therefore, the ground state is $2S_{\rm tot}+1$ fold degenerate.
    Clearly, the third component $S^{(3)}$ of spin is given by
    \begin{equation}
        S^{(3)}=\frac{1}{2}(2N_\uparrow -|\Lambda|)=N_\uparrow-\frac{1}{2}|\Lambda|.
    \end{equation}
    Therefore, when a ground state of the attractive Hubbard model with an even number of fermions
    is mapped to the ground state of the repulsive Hubbard model, $S^{(3)}$ must satisfy
    \begin{equation}
        -\frac{1}{2}\left||\Lambda_{\rm A}|-|\Lambda_{\rm B}|\right|\le S^{(3)}\le \frac{1}{2}\left||\Lambda_{\rm A}|-|\Lambda_{\rm B}|\right|.
    \end{equation}
    By combining these two relations, we have
    \begin{equation}
        |\Lambda_{\rm B}|\le N_\uparrow \le |\Lambda_{\rm A}|,
    \end{equation}
    where we have used $|\Lambda|=|\Lambda_{\rm A}|+|\Lambda_{\rm B}|$.
    In particular, the case of $|\Lambda_{\rm A}|>|\Lambda_{\rm B}|$ is important for superconductivity in the next section.

    \section{Long Range Order of Superconductivity}
    \label{Sec:LROsuper}

    {Throughout this section, we consider the Lieb lattice such that there exists a constant $0<a_0<1$ satisfying $|\Lambda_{\rm A}|-|\Lambda_{\rm B}|=a_0|\Lambda|$.}

    To begin with, consider the Hamiltonian $\check{H}^{(\Lambda)}=[U_{\rm S}^{(\Lambda)}]^\dagger H^{(\Lambda)}U_{\rm S}^{(\Lambda)}$
    of the repulsive Hubbard model. Similarly, we can define the expectation value by
    \begin{equation}
        \langle\!\langle\cdots\rangle\!\rangle_{\beta,\Lambda}^{\rm even}:=\frac{1}{\check{Z}_{\beta,\Lambda}^{\rm even}}
        \left[{\rm Tr}^{\rm e,e}(\cdots)e^{-\beta \check{H}^{(\Lambda)}}+{\rm Tr}^{\rm o,o}(\cdots)e^{-\beta \check{H}^{(\Lambda)}}\right]
    \end{equation}
    for the even number of fermions, $N=N_\uparrow+N_\downarrow= {\rm even}$, where
    \begin{equation}
        \check{Z}_{\beta,\Lambda}^{\rm even}:={\rm Tr}^{\rm e,e}e^{-\beta \check{H}^{(\Lambda)}}
        +{\rm Tr}^{\rm o,o}e^{-\beta \check{H}^{(\Lambda)}}.
    \end{equation}
    We assume that $|\Lambda|$ is an even integer. Then, the Shiba transformation does not change the even-oddness of
    the number $N_\downarrow$ of fermions with down spin $\downarrow$. Therefore, we have
    \begin{equation}
        \langle\cdots\rangle_{\beta,\Lambda}^{\rm even}
        =\langle\!\langle[U_{\rm S}^{(\Lambda)}]^\dagger H^{(\Lambda)}U_{\rm S}^{(\Lambda)}\rangle\!\rangle_{\beta,\Lambda}^{\rm even}.
    \end{equation}
    This yields
    \begin{eqnarray}
        & &\langle c_{x,\uparrow}^\dagger c_{x,\downarrow}^\dagger c_{y,\downarrow}c_{y,\uparrow}\rangle_{\beta,\Lambda}^{\rm even}
        +\langle c_{y,\uparrow}^\dagger c_{y,\downarrow}^\dagger c_{x,\downarrow}c_{x,\uparrow}\rangle_{\beta,\Lambda}^{\rm even}\nonumber\\
        &=&2\eta_x\eta_y[\langle\!\langle S_x^{(1)}S_y^{(1)}\rangle\!\rangle_{\beta,\Lambda}^{\rm even}
        +\langle\!\langle S_x^{(2)}S_y^{(2)}\rangle\!\rangle_{\beta,\Lambda}^{\rm even}
        +\delta_{x,y}\langle\!\langle S_x^{(3)}\rangle\!\rangle_{\beta,\Lambda}^{\rm even}]
    \end{eqnarray}
    in the same way as in (\ref{superspinexpc}). We also have
    \begin{equation}
        \langle c_{x,\uparrow}^\dagger c_{x,\downarrow}^\dagger c_{y,\downarrow}c_{y,\uparrow}\rangle_{\beta,\Lambda}^{\rm even}\ge 0.
    \end{equation}
    {From} these, one has
    \begin{equation}
        \langle [O_{\rm super}^{(\Lambda)}]^\dagger O_{\rm super}^{(\Lambda)}\rangle_{\beta,\Lambda}^{\rm even}\ge
        \frac{1}{|\Lambda|^2}\langle\!\langle (S^{(1)})^2+(S^{(2)})^2\rangle\!\rangle_{\beta,\Lambda}^{\rm even}
        +\frac{1}{|\Lambda|^2}\langle\!\langle S^{(3)}\rangle\!\rangle_{\beta,\Lambda}^{\rm even},
    \end{equation}
    where $S^{(i)}:=\sum_{x\in\Lambda}S_x^{(i)}$ for $i=1,2,3$. In the limit $\beta\nearrow\infty$, we obtain
    \begin{equation}
        \label{Eq.evenBD}
        \omega_{0,\Lambda}^{\rm even}( [O_{\rm super}^{(\Lambda)}]^\dagger O_{\rm super}^{(\Lambda)})\ge
        \frac{1}{|\Lambda|^2}\check{\omega}_{0,\Lambda}^{\rm even}((S^{(1)})^2+(S^{(2)})^2)
        +\frac{1}{|\Lambda|^2}\check{\omega}_{0,\Lambda}^{\rm even}(S^{(3)}),
    \end{equation}
    where
    \begin{equation}
        \check{\omega}_{0,\Lambda}^{\rm even}(\cdots)
        :=\lim_{\beta\nearrow\infty}\langle\!\langle\cdots\rangle\!\rangle_{\beta,\Lambda}^{\rm even}.
    \end{equation}
    As seen in the previous section, the ground state of the repulsive Hubbard model is $2S_{\rm tot}+1$-fold
    degenerate.
    {Since $|\Lambda|$ is even, we see that
        \[
        \check{\omega}_{0,\Lambda}^{\rm even}(S^{(3)})=
        \frac{1}{2S_{\rm tot}+1}\sum_{m=-S_{\rm tot}}^{S_{\rm tot}}m=0
        \]
        and
        \[
        \check{\omega}_{0,\Lambda}^{\rm even}((S^{(3)})^2)
        =
        \frac{1}{2S_{\rm tot}+1}\sum_{m=-S_{\rm tot}}^{S_{\rm tot}}m^2
        =
        \frac{1}{3}S_{\rm tot}(S_{\rm tot}+1).
        \]
        Using
        $
        (S^{(1)})^2+(S^{(2)})^2+(S^{(3)})^2
        =
        S_{\rm tot}(S_{\rm tot}+1)
        $
        on this sector, we obtain
        \[
        \check{\omega}_{0,\Lambda}^{\rm even}((S^{(1)})^2+(S^{(2)})^2)
        =
        \frac{2}{3}S_{\rm tot}(S_{\rm tot}+1).
        \]
        Substituting these identities into (\ref{Eq.evenBD}) gives}
    \begin{equation}
        \omega_{0,\Lambda}^{\rm even}( [O_{\rm super}^{(\Lambda)}]^\dagger O_{\rm super}^{(\Lambda)})\ge
        \frac{1}{|\Lambda|^2}\frac{2}{3}S_{\rm tot}(S_{\rm tot}+1)
    \end{equation}
    with $S_{\rm tot}=||\Lambda_{\rm A}|-|\Lambda_{\rm B}||/2$. By combining this with the bound (\ref{superLRObound}), we obtain

    \begin{theorem}
        \label{LROsuper}
        The following bound is valid:
        \begin{equation}
            \label{LROsuperbound}
            \omega_{0,\Lambda}([O_{\rm super}^{(\Lambda)}]^\dagger O_{\rm super}^{(\Lambda)})\ge
            \frac{1}{|\Lambda|^2}\frac{1}{3}S_{\rm tot}(S_{\rm tot}+1)
        \end{equation}
        for the superconducting long-range order. Therefore, when $S_{\rm tot}=||\Lambda_{\rm A}|-|\Lambda_{\rm B}||/2$ is of order
        of the volume $|\Lambda|$, the ground state $\omega_{0,\Lambda}(\cdots)$ shows the long-range order at half filling.
    \end{theorem}

    This theorem implies that {there exists a pure ground state with the non-vanishing superconducting order parameter} $O_{\rm super}^{(\Lambda)}$ in the infinite-volume limit~\cite{KomaTasaki}.
    In order to prove this statement, we write $H^{(\Lambda)}(B)$ for the Hamiltonian with a symmetry breaking field defined by
    \[
    H^{(\Lambda)}(B):=H^{(\Lambda)} - B|\Lambda| O_\Lambda \quad \mbox{with}\quad O_\Lambda:=O^{(\Lambda)}_\mathrm{super}
    +\left[O^{(\Lambda)}_\mathrm{super}\right]^\dagger
    \]
    and the external field $B>0$. The corresponding thermal expectation is
    \[
    \langle \cdots\rangle_{\beta, \Lambda}(B)
    :=\frac{1}{Z_{\beta,\Lambda}(B)}\mathrm{Tr} \left[(\cdots)e^{-\beta H^{(\Lambda)}(B)} \right]\quad
    \mbox{with} \quad Z_{\beta,\Lambda}(B):=\mathrm{Tr} \left[e^{-\beta H^{(\Lambda)}(B)} \right].
    \]
    Taking the zero temperature limit  $\beta \nearrow \infty$, we have
    \[
    \omega_{B,\Lambda}(\mathcal{A}):=\lim_{\beta \nearrow \infty}\langle \mathcal{A}\rangle_{\beta, \Lambda}(B)
    =\frac{1}{m}\sum_{j=1}^m \left\langle \Phi_j^{(\Lambda)}(B), \mathcal{A} \Phi_j^{(\Lambda)}(B)\right\rangle
    \]
    for any observable $\mathcal{A}$.
    Here, $\{\Phi_j^{(\Lambda)}(B)\}_{j=1}^m$ is an orthonormal basis of the $m$-fold degenerate ground states of $H^{(\Lambda)}(B)$.
    We define the unitary operator $U_\theta^{(\Lambda)}:=\exp(i \theta n^{(\Lambda)})$ for $\theta \in [0,2\pi)$.
    Since the Hamiltonian $H^{(\Lambda)}$ without the external field commutes with the total number $n^{(\Lambda)}$ of fermions,
    we can choose an orthonormal basis of the ground states of $H^{(\Lambda)}$ so that the basis $\{\Phi^{(\Lambda)}_j\}_{j=1}^\ell$
    consists of eigenstates of $n^{(\Lambda)}$, and thus of $U_\theta^{(\Lambda)}$ for all $\theta$.

    \begin{assumption}
        \label{assumpHop}
        For the hopping amplitudes $t_{x,y}$ of the Hamiltonian $H^{(\Lambda)}$ of (\ref{Ham}), there exists $0<t_0<+\infty$ such that
        \begin{equation}
            \label{Assumption}
            \sup_{x,y}|t_{x,y}|\le t_0.
        \end{equation}
    \end{assumption}

    Then, we have

    \begin{theorem}
        \label{sponsuper}
        Under Assumption~\ref{assumpHop}, it holds that
        \begin{equation}
            \label{sponSuper}
            \liminf_{B \searrow 0}\liminf_{|\Lambda| \nearrow \infty}\omega_{B,\Lambda}\bigl(O^{(\Lambda)}_\mathrm{super}
            +\left[O^{(\Lambda)}_\mathrm{super}\right]^\dagger\bigr)
            \ge q_0,
        \end{equation}
        where $q_0$ is a strictly positive constant which is independent of the volume $|\Lambda|$.
    \end{theorem}

    We stress that this result implies the existence of a pure state exhibiting {spontaneous symmetry breaking associated with the superconducting order parameter} in the infinite-volume limit.
    \bigskip

    \noindent
    \begin{proof}{Proof of Theorem~\ref{sponsuper}}
        The proof follows from Theorem~7.1 in Ref.~\cite{KomaTasaki}, {which is based on~\cite{HL,KHL}.}
        From the superconducting long-range order (\ref{LROsuperbound}), we may choose a ground state that exhibits the long-range order,
        namely, for each $\Lambda$ there exists $i$ such that
        \begin{equation}
            \label{ineq:LRO}
            \sigma_i^{(\Lambda)}:=
            \bigl\langle \Phi_i^{(\Lambda)}, \bigl(O^{(\Lambda)}_\mathrm{super}\left[O^{(\Lambda)}_\mathrm{super}\right]^\dagger
            + \left[O^{(\Lambda)}_\mathrm{super}\right]^\dagger O^{(\Lambda)}_\mathrm{super} \bigr)\Phi_i^{(\Lambda)}\bigr\rangle
            \ge
            \frac{2S_\mathrm{tot}\left(S_\mathrm{tot}+1\right)}{3 |\Lambda|^2}
            =:q_\Lambda^2,
        \end{equation}
        where we have used $\omega_{0,\Lambda}(O^{(\Lambda)}_\mathrm{super}[O^{(\Lambda)}_\mathrm{super}]^\dagger)
        =\omega_{0,\Lambda}([O^{(\Lambda)}_\mathrm{super}]^\dagger O^{(\Lambda)}_\mathrm{super})$
        that is obtained by the particle-hole transformation at half filling.
        Here we note that $q_0:=\liminf_{|\Lambda|\to\infty}q_\Lambda>0$, {since $|\Lambda_{\rm A}|-|\Lambda_{\rm B}|=a_0|\Lambda|$ for the Lieb lattices considered here.}
        Let $\Psi_i^{(\Lambda)}$ be a trial state given by
        \[
        \Psi_i^{(\Lambda)}:= \frac{1}{\sqrt{2}}\left(\Phi_i^{(\Lambda)}
        +\frac{O_\Lambda\Phi_i^{(\Lambda)}}{\bigl\|O_\Lambda\Phi_i^{(\Lambda)}\bigr\|}\right).
        \]
        Since $U_\theta^{(\Lambda)} c_{x,\sigma}^\dagger [U_\theta^{(\Lambda)}]^{\dagger}
        =e^{i\theta}c_{x,\sigma}^\dagger$, $U_\theta^{(\Lambda)} c_{x,\sigma} [U_\theta^{(\Lambda)}]^{\dagger}
        =e^{-i\theta}c_{x,\sigma}$, and $\{\Phi_j^{(\Lambda)}\}_j$ forms an orthonormal basis of eigenstates of $U_\theta^{(\Lambda)}$,
        it follows that for any positive integer $k$
        \begin{equation}
            \label{Equ.odd}
            \bigl\langle \Phi_i^{(\Lambda)},\left[O^{(\Lambda)}_\mathrm{super}\right]^k\Phi_i^{(\Lambda)} \bigr\rangle=0
            =\bigl\langle \Phi_i^{(\Lambda)},(\left[O^{(\Lambda)}_\mathrm{super}\right]^\dagger)^k\Phi_i^{(\Lambda)} \bigr\rangle
            \  \mbox{and} \
            \bigl\langle \Phi_i^{(\Lambda)},[O_\Lambda]^{2k+1}\Phi_i^{(\Lambda)} \bigr\rangle=0.
        \end{equation}
        This fact yields $\|\Psi_i^{(\Lambda)}\|=1$ and
        \begin{equation}
            \label{equ.LROeq}
            \begin{split}
                \bigl\langle \Psi_i^{(\Lambda)},O_\Lambda\Psi_i^{(\Lambda)} \bigr\rangle
                =\frac{\sigma_i^{(\Lambda)}}{\bigl\|O_\Lambda\Phi_i^{(\Lambda)}\bigr\|}
                =\sqrt{\sigma_i^{(\Lambda)}}
            \end{split}
        \end{equation}
        from the definition of $\sigma_i^{(\Lambda)}$ in (\ref{ineq:LRO}).
        For any ground state $\Phi_j^{(\Lambda)}(B)$, the variational principle implies that
        \[
        \bigl\langle \Phi_j^{(\Lambda)}(B), H^{(\Lambda)}(B)\Phi_j^{(\Lambda)}(B)\bigr\rangle
        \le
        \bigl\langle \Psi_i^{(\Lambda)}, H^{(\Lambda)}(B)\Psi_i^{(\Lambda)}\bigr\rangle.
        \]
        By a direct calculation, this inequality yields
        \begin{equation}
            \begin{split}
                \label{equ.variation}
                &\bigl\langle \Phi_j^{(\Lambda)}(B), O_\Lambda\Phi_j^{(\Lambda)}(B)\bigr\rangle\\
                &\ge \bigl\langle \Psi_i^{(\Lambda)}, O_\Lambda\Psi_i^{(\Lambda)}\bigr\rangle
                + \frac{1}{B|\Lambda|}\left[  	\bigl\langle \Phi_j^{(\Lambda)}(B), H^{(\Lambda)}\Phi_j^{(\Lambda)}(B)\bigr\rangle
                - \bigl\langle \Psi_i^{(\Lambda)}, H^{(\Lambda)}\Psi_i^{(\Lambda)}\bigr\rangle \right]\\
                &\ge
                \sqrt{\sigma_i^{(\Lambda)}}
                + \frac{1}{B|\Lambda|}\bigl[E_0^{(\Lambda)} - \bigl\langle \Psi_i^{(\Lambda)},
                H^{(\Lambda)}\Psi_i^{(\Lambda)}\bigr\rangle \bigr]\quad \mbox{for \ } B>0,
            \end{split}
        \end{equation}
        where we have used
        $\bigl\langle \Phi_j^{(\Lambda)}(B), H^{(\Lambda)}\Phi_j^{(\Lambda)}(B)\bigr\rangle \ge E_0^{(\Lambda)}$
        with the ground-state energy $E_0^{(\Lambda)}$ of $H^{(\Lambda)}$.
        For the second term in the last line, we deduce from the eigenvalue equation $H^{(\Lambda)} \Phi_i^{(\Lambda)}
        = E_0^{(\Lambda)}\Phi_i^{(\Lambda)}$ and \eqref{Equ.odd} that
        \begin{equation}
            \begin{split}
                \label{Equ.en}
                \bigl\langle \Psi_i^{(\Lambda)}, H^{(\Lambda)}\Psi_i^{(\Lambda)}\bigr\rangle-E_0^{(\Lambda)}
                &=
                \frac{ \bigl\langle\Phi_i^{(\Lambda)},O_\Lambda H^{(\Lambda)}O_\Lambda\Phi_i^{(\Lambda)}\bigr\rangle}{2\bigl\|O_\Lambda\Phi_i^{(\Lambda)}\bigr\|^2}-\frac{E_0^{(\Lambda)}}{2}\\
                &=
                \frac{2 \bigl\langle \Phi_i^{(\Lambda)},O_\Lambda H^{(\Lambda)}
                    O_\Lambda\Phi_i^{(\Lambda)}\bigr\rangle}{4\bigl\|O_\Lambda\Phi_i^{(\Lambda)}\bigr\|^2}
                - \frac{\bigl\langle \Phi_i^{(\Lambda)},(O^2_\Lambda H^{(\Lambda)}+H^{(\Lambda)}O^2_\Lambda) \Phi_i^{(\Lambda)}\bigr\rangle}
                {4\bigl\|O_\Lambda\Phi_i^{(\Lambda)}\bigr\|^2}\\
                &=
                \frac{\bigl\langle \Phi_i^{(\Lambda)}, \left[O_\Lambda , \left[H^{(\Lambda)},
                    O_\Lambda \right] \right] \Phi_i^{(\Lambda)}\bigr\rangle}
                {4\bigl\|O_\Lambda\Phi_i^{(\Lambda)}\bigr\|^2},
            \end{split}
        \end{equation}
        where we have used $[O_\Lambda, [H^{(\Lambda)}, O_\Lambda]]=2O_\Lambda H^{(\Lambda)} O_\Lambda -(O_\Lambda^2H^{(\Lambda)}
        +H^{(\Lambda)}O_\Lambda^2)$.
        Our next task is to evaluate the double commutator in \eqref{Equ.en}.
        Since $[n_{x,\uparrow}+n_{x,\downarrow}, c_{x,\downarrow} c_{x,\uparrow}]
        =-2c_{x,\downarrow} c_{x,\uparrow}$ and $[n_{x,\uparrow}n_{x,\downarrow}, c_{x,\downarrow} c_{x,\uparrow}]
        =-c_{x,\downarrow} c_{x,\uparrow}$, we have $[H^{(\Lambda)}_\mathrm{int}, O^{(\Lambda)}_\mathrm{super}
        +[ O^{(\Lambda)}_\mathrm{super}]^\dagger]=0$.
        Thus, it suffices to consider the hopping term.
        Using
        \[
        \left[c_{x,\uparrow}^\dagger c_{y,\uparrow}+c_{x,\downarrow}^\dagger c_{y,\downarrow},
        c_{x,\downarrow} c_{x,\uparrow}+c_{y,\downarrow} c_{y,\uparrow}\right]
        =-\left(c_{x,\downarrow} c_{y,\uparrow} +c_{y,\downarrow} c_{x,\uparrow}\right),
        \]
        together with its Hermitian conjugate, it follows that
        \[
        \sum_{\sigma=\uparrow,\downarrow}\sum_{x,y \in \Lambda}t_{x,y}\left[c_{x,\sigma}^\dagger c_{y,\sigma}
        + c_{y,\sigma}^\dagger c_{x,\sigma}, O_\Lambda \right]
        =
        -\frac{2}{|\Lambda|}
        \sum_{x,y \in \Lambda}(t_{x,y}+t_{y,x}) \left(c_{x,\downarrow} c_{y,\uparrow} -c_{y,\uparrow}^\dagger c_{x,\downarrow}^\dagger \right).
        \]
        Similarly, we find
        \[
        \left[c_{x,\downarrow} c_{x,\uparrow}+c_{y,\downarrow} c_{y,\uparrow},c_{x,\downarrow} c_{y,\uparrow}
        -c_{y,\uparrow}^\dagger c_{x,\downarrow}^\dagger \right]
        =c_{y,\uparrow}^\dagger c_{x,\uparrow}+c_{x,\downarrow}^\dagger c_{y,\downarrow},
        \]
        and hence
        \[
        \left[O^{(\Lambda)}_\mathrm{super}+\left[O^{(\Lambda)}_\mathrm{super}\right]^\dagger,
        \left[H^{(\Lambda)}_{\mathrm{hop}}, O^{(\Lambda)}_\mathrm{super}+\left[O^{(\Lambda)}_\mathrm{super}\right]^\dagger\right] \right]
        =-\frac{4}{|\Lambda|^2}H^{(\Lambda)}_{\mathrm{hop}}.
        \]
        {From} the assumption \eqref{Assumption}, we have
        \[
        \bigl|\bigl\langle \Phi_i^{(\Lambda)}, \bigl[O_\Lambda, \bigl[H^{(\Lambda)}_{\mathrm{hop}},
        O_\Lambda\bigr] \bigr] \Phi_i^{(\Lambda)}\bigr\rangle\bigr|
        \le
        \frac{8t_0}{|\Lambda|}
        \]
        Combining this with \eqref{equ.variation} and \eqref{Equ.en}, we arrive at
        \begin{equation}
            \begin{split}
                \label{Equ.jGS}
                \bigl\langle \Phi_j^{(\Lambda)}(B), O_\Lambda\Phi_j^{(\Lambda)}(B)\bigr\rangle
                &\ge
                \sqrt{\sigma_i^{(\Lambda)}}
                -\frac{2t_0}{B\left|\Lambda \right|^2\bigl\|O_\Lambda\Phi_i^{(\Lambda)}\bigr\|^2} \\
                &\ge
                q_\Lambda
                -\frac{2t_0}{B\left|\Lambda \right|^2 q_\Lambda^2},
            \end{split}
        \end{equation}
        where we have used the lower bound on the long-range order parameter \eqref{ineq:LRO}.
        By construction, the inequality \eqref{Equ.jGS} holds for every ground state $\Phi_j^{(\Lambda)}(B)$.
        Therefore, we obtain
        \begin{equation}
            \omega_{B,\Lambda}\bigl(O^{(\Lambda)}_\mathrm{super} +[O^{(\Lambda)}_\mathrm{super}]^\dagger\bigr)
            \ge
            q_\Lambda
            -\frac{2t_0}{B\left|\Lambda \right|^2 q_\Lambda^2}.
        \end{equation}
        Taking the infinite volume limit $|\Lambda| \nearrow \infty$ and then letting $B \searrow 0$,
        we obtain the desired bound (\ref{sponSuper}).
    \end{proof}

    \section{Spectral Gap of Single-Fermion Excitations}
    \label{Sec:GapSFE}
    \subsection{The Case of Hypercubic Lattice}
    From now on, we will deal with excitations of the Hubbard model.
    In this section, we require lattice translational invariance for the Hamiltonian $H^{(\Lambda)}$.
    More precisely, we first consider the usual hypercubic lattice $\ze^d$ of dimension $d=1,2,3,\ldots$,
    and then we will discuss the case of Lieb lattice.
    Therefore, for all the non-vanishing hopping amplitudes,
    we choose $t_{x,y}=-t$ with a positive constant $t$, and for the coupling constants of the interactions, $U_x=U$
    at all the lattice sites $x\in\Lambda\subset\ze^d$ with a positive constant $U$.
    Then, the translationally invariant Hamiltonian $H^{(\Lambda)}$ is written
    \begin{equation}
        \label{Hamtrans}
        H^{(\Lambda)}=-t\sum_{\sigma\in\{\uparrow,\downarrow\}}\sum_{\substack{x,y\in\Lambda \\|x-y|=1}}
        c_{x,\sigma}^\dagger c_{y,\sigma}
        -U\sum_{x\in\Lambda}\left(n_{x,\uparrow}-\frac{1}{2}\right)\left(n_{x,\downarrow}-\frac{1}{2}\right)
        +\mu\sum_{\sigma\in\{\uparrow,\downarrow\}}\sum_{x\in\Lambda}n_{x,\sigma}.
    \end{equation}
    Here, we have assumed the nearest neighbour hopping for simplicity, and introduced the chemical potential $\mu\ge 0$ to realize
    a low density of fermions below.

    In order to construct an excitation above a ground state $\Phi_{\rm GS}$,
    we introduce a fermionic operator with up spin $\uparrow$,
    \begin{equation}
        \label{calAup}
        \mathcal{A}_\uparrow:=\sum_{x\in\Lambda}\alpha_x c_{x,\uparrow}^\dagger,
    \end{equation}
    where $\alpha_x\in \co$, and consider the excited state $\mathcal{A}_\uparrow \Phi_{\rm GS}$ above the ground state $\Phi_{\rm GS}$.
    Note that
    \begin{eqnarray}
        \mathcal{A}_\uparrow \mathcal{A}_\uparrow^\dagger + \mathcal{A}_\uparrow^\dagger \mathcal{A}_\uparrow
        &=&\sum_{x,y\in\Lambda}\alpha_x\alpha_y^\ast c_{x,\uparrow}^\dagger c_{y,\uparrow}
        +\sum_{x,y\in\Lambda}\alpha_x^\ast \alpha_y c_{x,\uparrow} c_{y,\uparrow}^\dagger\nonumber\\
        &=&\sum_{x,y\in\Lambda}\alpha_x\alpha_y^\ast c_{x,\uparrow}^\dagger c_{y,\uparrow}
        +\sum_{x,y\in\Lambda}\alpha_y^\ast \alpha_x c_{y,\uparrow} c_{x,\uparrow}^\dagger\nonumber\\
        &=&\sum_{x,y\in\Lambda} \alpha_x\alpha_y^\ast (c_{x,\uparrow}^\dagger c_{y,\uparrow}+c_{y,\uparrow} c_{x,\uparrow}^\dagger)
        =\sum_{x\in\Lambda} |\alpha_x|^2,
        \label{AAbound}
    \end{eqnarray}
    where we have used the anti-commutation relation,
    $c_{x,\uparrow}^\dagger c_{y,\uparrow}+c_{y,\uparrow} c_{x,\uparrow}^\dagger=\delta_{x,y}$. By relying on this relation,
    we choose the normalization to be
    \begin{equation}
        \label{normalpha}
        \sum_{x\in\Lambda} |\alpha_x|^2=1.
    \end{equation}
    Therefore, we have $\mathcal{A}_\uparrow \mathcal{A}_\uparrow^\dagger + \mathcal{A}_\uparrow^\dagger \mathcal{A}_\uparrow=1$.
    From this relation, {we restrict our attention to single-fermion excitations $\mathcal{A}_\uparrow\Phi_{\rm GS}$ satisfying}
    \begin{equation}
        \label{assumAnorm}
        \langle\Phi_{\rm GS},\mathcal{A}_\uparrow^\dagger \mathcal{A}_\uparrow\Phi_{\rm GS}\rangle\ge \frac{1}{2}\Vert\Phi_{\rm GS}\Vert^2.
    \end{equation}
    {As we will see in Example~\ref{Ex.exc}, such an excitation can be constructed in the low-density regime.}
    Clearly, this relation does not hold when all the lattice sites are already occupied by fermions with up spin in a ground state $\Phi_{\rm GS}$,
    if it exists. But we will consider the regime of low density of fermions with up spin.
    Actually, we can expect $\Vert \mathcal{A}_\uparrow\Phi_{\rm GS}\Vert\approx 1$ in the regime,
    while $\Vert \mathcal{A}_\uparrow\Phi_{\rm GS}\Vert\approx 0$ is expected in the high density regime.

    {Now we define the excitation energy associated with the state $\mathcal{A}_\uparrow\Phi_{\rm GS}$ by}
    \[
    \Delta(\mathcal{A}_\uparrow\Phi_{\rm GS}):=\frac{\|(H^{(\Lambda)}-E_{\rm GS})\mathcal{A}_\uparrow\Phi_{\rm GS}\|}{\|\mathcal{A}_\uparrow\Phi_{\rm GS}\|}.
    \]

    Our result in this subsection is as follows:

    \begin{theorem}
        \label{thmGap}
        {Let $\Phi_{\rm GS}$ be the ground state of the translationally invariant
            Hamiltonian $H^{(\Lambda)}$ of (\ref{Hamtrans}) on the lattice $\Lambda\subset \ze^d$ with an even number $N>0$ of fermions, and let $\mathcal{A}_\uparrow$ be given by (\ref{calAup}).
            Assume that the filling factor $\nu_\downarrow:=N_\downarrow/|\Lambda|$ for the number $N_\downarrow$ of
            fermions with down spin satisfies $\nu_\downarrow<1/8$. Then, for a sufficiently small $|t|/U$, the excitation energy $\Delta(\mathcal{A}_\uparrow\Phi_{\rm GS})$ is bounded from below by a positive constant
            if the state $\mathcal{A}_\uparrow\Phi_{\rm GS}$ satisfies
            the norm condition (\ref{assumAnorm}).}
    \end{theorem}

    {In other words, adding a single fermion costs a strictly positive energy above the ground state.
        In this paper, we also refer to $\Delta(\mathcal{A}_\uparrow\Phi_{\rm GS})$ as a spectral gap of single-fermion excitations. }

    {As in the following Example~\ref{Ex.exc}, we may take $\mathcal{A}_\uparrow=c^\dagger_{x_0,\uparrow}$ so that \eqref{assumAnorm} holds in the low-density regime.
        Following Lieb's theorem \cite{Lieb}, for any even number $N>0$ of the fermions, the ground state is unique and shows spin singlet.
        Therefore, the number $N_\uparrow$ of fermions with up spin satisfies $N_\uparrow=N_\downarrow$ and $N=N_\uparrow+N_\downarrow$.
        Clearly, the filling condition $\nu_\downarrow<1/8$ can be realized.}

    \begin{example}
        \label{Ex.exc}
        The simplest example is given by
        \begin{equation}
            \mathcal{A}_\uparrow=c_{x_0,\uparrow}^\dagger
        \end{equation}
        for a fixed site $x_0\in\Lambda$. Clearly, one has
        \begin{eqnarray}
            \langle\Phi_{\rm GS},\mathcal{A}_\uparrow^\dagger \mathcal{A}_\uparrow\Phi_{\rm GS}\rangle
            &=&\langle\Phi_{\rm GS},c_{x_0,\uparrow}c_{x_0,\uparrow}^\dagger\Phi_{\rm GS}\rangle \nonumber\\
            &=&\Vert \Phi_{\rm GS}\Vert^2-\langle \Phi_{\rm GS},c_{x_0,\uparrow}^\dagger c_{x_0,\uparrow}\Phi_{\rm GS}\rangle\\
            &=&\Vert \Phi_{\rm GS}\Vert^2(1-\nu_\uparrow),
        \end{eqnarray}
        where $\nu_\uparrow$ is the density of fermions with up spin, i.e., $\nu_\uparrow=N_\uparrow/|\Lambda|$, and we have used
        the translational invariance of the ground state $\Phi_{\rm GS}$. Thus, this example satisfies the condition (\ref{assumAnorm})
        in the low-density regime $\nu_\uparrow \ll 1$. Next, consider
        \begin{equation}
            \mathcal{A}_\uparrow=\alpha_{x_0}c_{x_0,\uparrow}^\dagger+\alpha_{y_0}c_{y_0,\uparrow}^\dagger
        \end{equation}
        with fixed two sites $x_0, y_0\in\Lambda$ $(x_0\ne y_0)$. By using the normalization (\ref{normalpha}), one has
        \begin{equation}
            \mathcal{A}_\uparrow^\dagger \mathcal{A}_\uparrow=1-|\alpha_{x_0}|^2n_{x_0,\uparrow}-|\alpha_{y_0}|^2n_{y_0,\uparrow}
            +\alpha_{x_0}^\ast \alpha_{y_0}c_{x_0,\uparrow}c_{y_0,\uparrow}^\dagger+\alpha_{x_0}\alpha_{y_0}^\ast c_{y_0,\uparrow}
            c_{x_0,\uparrow}^\dagger.
        \end{equation}
        The first three terms on the right-hand side yield $1-\nu_\uparrow$ in the same way. Therefore, it is enough to estimate
        the last two terms. Note that
        \begin{eqnarray}
            & &\alpha_{x_0}^\ast \alpha_{y_0}c_{x_0,\uparrow}c_{y_0,\uparrow}^\dagger+\alpha_{x_0}\alpha_{y_0}^\ast c_{y_0,\uparrow}
            c_{x_0,\uparrow}^\dagger \nonumber\\
            &=&|\alpha_{x_0}||\alpha_{y_0}|
            [(c_{x_0,\uparrow}c_{y_0,\uparrow}^\dagger+c_{y_0,\uparrow}c_{x_0,\uparrow}^\dagger)\cos \theta
            +i(c_{x_0,\uparrow}c_{y_0,\uparrow}^\dagger-c_{y_0,\uparrow}c_{x_0,\uparrow}^\dagger)\sin \theta]
        \end{eqnarray}
        with $\theta\in[0,2\pi)$. The right-hand side can be estimated by using the following inequalities:
        \begin{equation}
            (c_{x_0,\uparrow}^\dagger \pm c_{y_0,\uparrow}^\dagger)(c_{x_0,\uparrow}\pm c_{y_0,\uparrow})\ge 0
            \ \ \mbox{and}  \ \ (c_{x_0,\uparrow}^\dagger \pm i c_{y_0,\uparrow}^\dagger)(c_{x_0,\uparrow}\mp i c_{y_0,\uparrow})\ge 0.
        \end{equation}
        Consequently, we have the desired result,
        \begin{equation}
            \langle\Phi_{\rm GS},\mathcal{A}_\uparrow^\dagger \mathcal{A}_\uparrow\Phi_{\rm GS}\rangle\ge
            \Vert \Phi_{\rm GS}\Vert^2(1-3\nu_\uparrow),
        \end{equation}
        where we have also used $|\alpha_{x_0}||\alpha_{y_0}|\le (|\alpha_{x_0}|^2+|\alpha_{y_0}|^2)/2=1/2$.
        The case with three or more sites is straightforward. Clearly, in order to lower the hopping energy for the excitation,
        it is necessary to increase the number of sites. However, the number of sites is not intrinsic for the existence of a gap
        because the contribution from the hopping energy is always of order $|t|$ as we will see below.
    \end{example}

    Thus the class of the single-fermion excitations that satisfies the condition (\ref{assumAnorm}) is well-defined
    for the low-density regime of fermions. In addition, in the infinite-volume formalism, the excitations above a ground state
    are constructed from their finite supports \cite{Koma1}.

    \begin{proof}{Proof of Theorem~\ref{thmGap}}
        In order to prove the existence of a spectral gap for single-fermionic excitations above the ground state $\Phi_{\rm GS}$,
        we consider
        \begin{equation}
            \label{HcommuA}
            [H^{(\Lambda)},\mathcal{A}_\uparrow]\Phi_{\rm GS}=(H^{(\Lambda)}\mathcal{A}_\uparrow-\mathcal{A}_\uparrow H^{(\Lambda)})\Phi_{\rm GS}
            =(H^{(\Lambda)}-E_{\rm GS})\mathcal{A}_\uparrow\Phi_{\rm GS},
        \end{equation}
        where $E_{\rm GS}$ is the energy of the ground state $\Phi_{\rm GS}$. We have

        \begin{lemma}
            Under the condition~(\ref{assumAnorm}), the following bound is valid:
            \begin{equation}
                \label{HEbound}
                \frac{\Vert(H^{(\Lambda)}-E_{\rm GS})\mathcal{A}_\uparrow\Phi_{\rm GS}\Vert}{\Vert \mathcal{A}_\uparrow\Phi_{\rm GS}\Vert}\ge
                \Bigl(\frac{U}{2}+\mu\Bigr)-{\sqrt{2}}{|t|}\frac{\Vert \tilde{\mathcal{A}}_\uparrow\Phi_{\rm GS}\Vert}{\Vert\Phi_{\rm GS}\Vert}
                -\sqrt{2}U\frac{\Vert \mathcal{A}_{\uparrow,\downarrow}\Phi_{\rm GS}\Vert}{\Vert\Phi_{\rm GS}\Vert},
            \end{equation}
            where we have written
            \begin{equation}
                \label{calAtilde}
                \tilde{\mathcal{A}}_\uparrow:=\sum_{x\in\Lambda}\tilde{\alpha}_x c_{x,\uparrow}^\dagger\quad
                \mbox{with}\quad \tilde{\alpha}_x:=\sum_{y\in\Lambda:|x-y|=1}\alpha_y,
            \end{equation}
            and
            \begin{equation}
                \mathcal{A}_{\uparrow,\downarrow}:=\sum_{x\in\Lambda}\alpha_x c_{x,\uparrow}^\dagger n_{x,\downarrow}.
            \end{equation}
        \end{lemma}

        \begin{proof}{Proof of Lemma}
            The commutator on the left-hand side of (\ref{HcommuA}) is computed as follows:
            \begin{eqnarray}
                & &[H^{(\Lambda)},\mathcal{A}_\uparrow]\nonumber\\
                &=&-t\sum_{x\in\Lambda}\alpha_x\sum_{y\in\Lambda}\sum_{z\in\Lambda:|z-y|=1}
                [c_{y,\uparrow}^\dagger c_{z,\uparrow},c_{x,\uparrow}^\dagger]
                -U\sum_{x\in\Lambda}\alpha_x[(n_{x,\uparrow}-1/2),c_{x,\uparrow}^\dagger](n_{x,\downarrow}-1/2)+\mu\mathcal{A}_\uparrow
                \nonumber\\ &=&
                -t\sum_{x\in\Lambda}\alpha_x\sum_{y\in\Lambda:|x-y|=1}
                [c_{y,\uparrow}^\dagger c_{x,\uparrow},c_{x,\uparrow}^\dagger]
                -U\sum_{x\in\Lambda}\alpha_x c_{x,\uparrow}^\dagger(n_{x,\downarrow}-1/2)+\mu\mathcal{A}_\uparrow\nonumber\\
                &=&-t\sum_{x\in\Lambda}\alpha_x\sum_{y\in\Lambda:|x-y|=1}c_{y,\uparrow}^\dagger
                -U\sum_{x\in\Lambda}\alpha_x c_{x,\uparrow}^\dagger(n_{x,\downarrow}-1/2)+\mu\mathcal{A}_\uparrow\nonumber\\
                &=&-t\sum_{y\in\Lambda}\tilde{\alpha}_y c_{y,\uparrow}^\dagger
                -U\sum_{x\in\Lambda}\alpha_x c_{x,\uparrow}^\dagger(n_{x,\downarrow}-1/2)+\mu\mathcal{A}_\uparrow\nonumber\\
                &=&\Bigl(\frac{U}{2}+\mu\Bigr)\mathcal{A}_\uparrow-t\sum_{y\in\Lambda}\tilde{\alpha}_y c_{y,\uparrow}^\dagger
                -U\sum_{x\in\Lambda}\alpha_x c_{x,\uparrow}^\dagger n_{x,\downarrow},
            \end{eqnarray}
            where we have used $[c_{y,\uparrow}^\dagger c_{x,\uparrow},c_{x,\uparrow}^\dagger]=c_{y,\uparrow}^\dagger$ for $y\ne x$.
            Therefore, we have
            \begin{equation}
                \Vert(H^{(\Lambda)}-E_{\rm GS})\mathcal{A}_\uparrow\Phi_{\rm GS}\Vert\ge
                \Bigl(\frac{U}{2}+\mu\Bigr)\Vert \mathcal{A}_\uparrow\Phi_{\rm GS}\Vert-|t|\Vert \tilde{\mathcal{A}}_\uparrow\Phi_{\rm GS}\Vert
                -U\Vert \mathcal{A}_{\uparrow,\downarrow}\Phi_{\rm GS}\Vert,
            \end{equation}
            where we have used the inequality $\Vert\Phi_1+\Phi_2\Vert\ge \Vert \Phi_1\Vert -\Vert \Phi_2\Vert$
            for any two states $\Phi_1$ and $\Phi_2$ in the Fock space.\footnote{The inequality
                is derived from the well-known triangle inequality $\Vert\Phi_1+\Phi_2\Vert\le \Vert \Phi_1\Vert+\Vert\Phi_2\Vert$.}
            Further, by using the assumption (\ref{assumAnorm}), one can obtain the bound (\ref{HEbound}).
        \end{proof}

        Since the bound (\ref{Zboundeven}) holds also when introducing the chemical potential $\mu$, one has
        \begin{equation}
            \frac{1}{\beta}\log {\rm Tr}(e^{-\beta H^{(\Lambda)}})\le \frac{1}\beta \log 2
            +\frac{1}{\beta}\log \left[{\rm Tr}^{\rm e,e}(e^{-\beta H^{(\Lambda)}})+{\rm Tr}^{\rm o,o}(e^{-\beta H^{(\Lambda)}})\right]
        \end{equation}
        Therefore, in the limit $\beta\nearrow \infty$, we have
        \begin{equation}
            E_{\rm GS}\ge \min \{E_{\rm GS}^{\rm e,e},E_{\rm GS}^{\rm o,o}\},
        \end{equation}
        where $E_{\rm GS}^{\rm p,p}$ is the ground state energy with the number $N_\sigma={\rm p}$ of fermions
        with spin $\sigma\in\{\uparrow,\downarrow\}$ and ${\rm p}\in\{{\rm even, odd}\}$.\footnote{A similar bound
            was obtained for the ground-state energy of interacting spinless fermions in Eq.~(17) of Ref.~\cite{WHXX} and of the SU($N$) Hubbard model in Lemma~12 of Ref.~\cite{YK}.
            See also \cite{Tian,TT}.} Consequently,
        the ground-state energy $E_{\rm GS}$ is equal to the energy of the ground state with an even number of fermions,
        irrespective of the value of the chemical potential $\mu$.
        In other words, the energy of the ground state with an odd number of
        fermions is always equal to or greater than that with an even number of fermions for any given chemical potential $\mu$.
        Thus, to prove the existence of a spectral gap of the single-fermion excitation above the ground state $\Phi_{\rm GS}$ with an even number of fermions,
        it is enough to estimate the second and third terms on the right-hand side of (\ref{HEbound}).

        Consider first the second term on the right-hand side of (\ref{HEbound}). In the same way as in (\ref{AAbound}), we have
        \begin{equation}
            \label{expGSAAbound}
            \frac{\langle \Phi_{\rm GS},\tilde{\mathcal A}_\uparrow^\dagger \tilde{\mathcal A}_\uparrow\Phi_{\rm GS}\rangle}{\Vert \Phi_{\rm GS}\Vert^2}\le
            \frac{\langle \Phi_{\rm GS},\tilde{\mathcal A}_\uparrow^\dagger \tilde{\mathcal A}_\uparrow\Phi_{\rm GS}\rangle}{\Vert \Phi_{\rm GS}\Vert^2}+
            \frac{\langle \Phi_{\rm GS},\tilde{\mathcal A}_\uparrow \tilde{\mathcal A}_\uparrow^\dagger\Phi_{\rm GS}\rangle}{\Vert \Phi_{\rm GS}\Vert^2}
            = \sum_{x\in\Lambda}|\tilde{\alpha}_x|^2.
        \end{equation}
        From the definition of $\tilde{\alpha}_y$ in (\ref{calAtilde}), one has
        \begin{equation}
            |\tilde{\alpha}_y|^2\le \gamma_{\rm nn}\sum_{x\in\Lambda:|x-y|=1}|\alpha_x|^2
        \end{equation}
        with a positive constant $\gamma_{\rm nn}$ which is determined by the number of the nearest neighbour sites.
        Therefore, the right-hand side of (\ref{expGSAAbound}) is bounded by a positive constant $\tilde{\gamma}_{\rm nn}$ as follows:
        \begin{equation}
            \sum_{x\in\Lambda}|\tilde{\alpha}_x|^2\le \tilde{\gamma}_{\rm nn},
        \end{equation}
        where we have used the normalization (\ref{normalpha}) of $\alpha_x$. In consequence, we have
        \begin{equation}
            \label{tildeAupbound}
            \frac{\Vert \tilde{\mathcal{A}}_\uparrow\Phi_{\rm GS}\Vert}{\Vert\Phi_{\rm GS}\Vert}\le \sqrt{\tilde{\gamma}_{\rm nn}}.
        \end{equation}

        Similarly, for the third term on the right-hand side of (\ref{HEbound}), we have
        \begin{eqnarray}
            \langle\Phi_{\rm GS},\mathcal{A}_{\uparrow,\downarrow}^\dagger\mathcal{A}_{\uparrow,\downarrow}\Phi_{\rm GS}\rangle
            &\le& \langle\Phi_{\rm GS},\mathcal{A}_{\uparrow,\downarrow}^\dagger\mathcal{A}_{\uparrow,\downarrow}\Phi_{\rm GS}\rangle
            +\langle\Phi_{\rm GS},\mathcal{A}_{\uparrow,\downarrow}\mathcal{A}_{\uparrow,\downarrow}^\dagger\Phi_{\rm GS}\rangle\nonumber\\
            &=& \sum_{x\in\Lambda}|\alpha_x|^2\langle\Phi_{\rm GS},n_{x,\downarrow}\Phi_{\rm GS}\rangle\nonumber\\
            &=&\frac{1}{|\Lambda|}\sum_{x\in\Lambda}\langle\Phi_{\rm GS},n_{x,\downarrow}\Phi_{\rm GS}\rangle
            \sum_{x\in\Lambda}|\alpha_x|^2\nonumber\\
            &=&\frac{1}{|\Lambda|}\sum_{x\in\Lambda}\langle\Phi_{\rm GS},n_{x,\downarrow}\Phi_{\rm GS}\rangle
            =\frac{N_\downarrow}{|\Lambda|}\Vert\Phi_{\rm GS}\Vert^2,
        \end{eqnarray}
        where we have used the translational invariance of the ground state $\Phi_{\rm GS}$ and the normalization (\ref{normalpha}),
        and written $N_\downarrow$ for the number of the fermions with down spin in the ground state $\Phi_{\rm GS}$.
        By using this and the bound (\ref{tildeAupbound}) on the right-hand side of (\ref{HEbound}), we obtain
        \begin{equation}
            \frac{\Vert(H^{(\Lambda)}-E_{\rm GS})\mathcal{A}_\uparrow\Phi_{\rm GS}\Vert}{\Vert \mathcal{A}_\uparrow\Phi_{\rm GS}\Vert}\ge
            \Bigl(\frac{U}{2}+\mu\Bigr)-{\sqrt{2\tilde{\gamma}_{\rm nn}}}{|t|}-\sqrt{2\nu_\downarrow}U,
        \end{equation}
        where we have written $\nu_\downarrow:=N_\downarrow/|\Lambda|$. Therefore, if the filling factor $\nu_\downarrow$ of
        the ground state $\Phi_{\rm GS}$ satisfies $\nu_\downarrow<1/8$, then the excitation $\mathcal{A}_\uparrow\Phi_{\rm GS}$
        has a spectral gap above the ground state for a sufficiently small $|t|/U$.
    \end{proof}
    \subsection{The Case of Lieb Lattice}
    \label{Sec:LiebLatticeGap}

    Let us consider the case of a Lieb lattice which consists of the two sublattices $\Lambda_{\rm A}$
    and $\Lambda_{\rm B}$. We still require translational invariance for the Hamiltonian $H^{(\Lambda)}$.
    We also assume a site symmetry that yields the uniform density of fermions on both $\Lambda_{\rm A}$ and $\Lambda_{\rm B}$
    in the translationally invariant ground state.
    For simplicity, we assume that the Hamiltonian $H^{(\Lambda)}$ can be written in the same form as (\ref{Hamtrans}).
    We write
    \begin{equation}
        \label{LieblatticeCond}
        \frac{|\Lambda_{\rm A}|}{|\Lambda|}=1-a \quad \mbox{and}\quad \frac{|\Lambda_{\rm B}|}{|\Lambda|}=a\quad \mbox{with} \ \ 0<a<1/2.
    \end{equation}
    We obtain

    \begin{theorem}
        \label{thmGapLieb}
        {Let $\Phi_{\rm GS}$ be the ground state of the translationally invariant Hamiltonian
            $H^{(\Lambda)}$ of~\eqref{Hamtrans} on the Lieb lattice $\Lambda$ with an even number $N>0$ of fermions, and let $\mathcal{A}_\uparrow$ be given by (\ref{calAup}).
            Here $\Lambda$ has the above site symmetry and satisfies~\eqref{LieblatticeCond} with $0<a<1/2$.
            Assume that the filling factor $\nu_\downarrow:=N_\downarrow/|\Lambda|$ satisfies $\nu_\downarrow<a/8$.
            Then, for a sufficiently small $|t|/U$, the excitation energy $\Delta(\mathcal{A}_\uparrow\Phi_{\rm GS})$  is bounded from below by a positive constant,
            if the state $\mathcal{A}_\uparrow\Phi_{\rm GS}$
            satisfies the norm condition~\eqref{assumAnorm}.}
    \end{theorem}

    \noindent
    Namely, in comparison with the statement of Theorem~\ref{thmGap},
    the condition for the existence of a non-vanishing spectral gap is replaced by
    $\nu_\downarrow<{a}/{8}$ about the filling factor $\nu_\downarrow$ in the case of the Lieb lattice.
    \bigskip

    \begin{proof}{Proof of Theorem~\ref{thmGapLieb}}
        One notices that the arguments to prove the existence of a spectral gap are the same as those in the preceding section
        except for the estimate of the third term on the right-hand side of (\ref{HEbound}).
        In the same way as in the proof of Theorem~\ref{thmGap},  for the term, we have
        \begin{eqnarray}
            \langle\Phi_{\rm GS},\mathcal{A}_{\uparrow,\downarrow}^\dagger\mathcal{A}_{\uparrow,\downarrow}\Phi_{\rm GS}\rangle
            &\le& \langle\Phi_{\rm GS},\mathcal{A}_{\uparrow,\downarrow}^\dagger\mathcal{A}_{\uparrow,\downarrow}\Phi_{\rm GS}\rangle
            +\langle\Phi_{\rm GS},\mathcal{A}_{\uparrow,\downarrow}\mathcal{A}_{\uparrow,\downarrow}^\dagger\Phi_{\rm GS}\rangle\nonumber\\
            &=& \sum_{x\in\Lambda}|\alpha_x|^2\langle\Phi_{\rm GS},n_{x,\downarrow}\Phi_{\rm GS}\rangle\nonumber\\
            &=&\sum_{x\in\Lambda_{\rm A}}|\alpha_x|^2\langle\Phi_{\rm GS},n_{x,\downarrow}\Phi_{\rm GS}\rangle
            +\sum_{x\in\Lambda_{\rm B}}|\alpha_x|^2\langle\Phi_{\rm GS},n_{x,\downarrow}\Phi_{\rm GS}\rangle\nonumber\\
            &=&\frac{N_{\downarrow,{\rm A}}}{|\Lambda_{\rm A}|}\sum_{x\in\Lambda_{\rm A}}|\alpha_x|^2\Vert\Phi_{\rm GS}\Vert^2
            +\frac{N_{\downarrow,{\rm B}}}{|\Lambda_{\rm B}|}\sum_{x\in\Lambda_{\rm B}}|\alpha_x|^2\Vert\Phi_{\rm GS}\Vert^2\nonumber\\
            &\le& \left[\frac{N_{\downarrow,{\rm A}}}{|\Lambda_{\rm A}|}
            +\frac{N_{\downarrow,{\rm B}}}{|\Lambda_{\rm B}|}\right]\Vert\Phi_{\rm GS}\Vert^2,
        \end{eqnarray}
        where we have written
        \begin{equation}
            N_{\downarrow,{\rm A}}=\sum_{x\in\Lambda_{\rm A}}
            \frac{\langle\Phi_{\rm GS},n_{x,\downarrow}\Phi_{\rm GS}\rangle}{\Vert\Phi_{\rm GS}\Vert^2}
            \quad \mbox{and}\quad
            N_{\downarrow,{\rm B}}=\sum_{x\in\Lambda_{\rm B}}
            \frac{\langle\Phi_{\rm GS},n_{x,\downarrow}\Phi_{\rm GS}\rangle}{\Vert\Phi_{\rm GS}\Vert^2},
        \end{equation}
        and we have used the translational invariance and the assumption of the site symmetry
        that yield the uniform density of fermions on both $\Lambda_{\rm A}$ and $\Lambda_{\rm B}$.
        In order to evaluate the above right-hand side, we use the condition (\ref{LieblatticeCond}) for the Lieb lattice.
        These yield
        \begin{equation}
            \frac{N_{\downarrow,{\rm A}}}{|\Lambda_{\rm A}|}+\frac{N_{\downarrow,{\rm B}}}{|\Lambda_{\rm B}|}
            =\frac{N_{\downarrow,{\rm A}}}{(1-a)|\Lambda|}+\frac{N_{\downarrow,{\rm B}}}{a|\Lambda|}\le \frac{N_\downarrow}{a|\Lambda|}
            =\frac{\nu_\downarrow}{a}.
        \end{equation}
        Therefore, we have
        \begin{equation}
            \frac{\Vert\mathcal{A}_{\uparrow,\downarrow}\Phi_{\rm GS}\Vert}{\Vert\Phi_{\rm GS}\Vert}\le \sqrt{\frac{\nu_\downarrow}{a}}.
        \end{equation}
        This implies that the condition for the existence of a non-vanishing spectral gap is replaced by
        $\nu_\downarrow<{a}/{8}$ for the filling factor $\nu_\downarrow$ in the case of the Lieb lattice.
    \end{proof}

    \section{Low-Lying Excitations}
    \label{Sec:LLE}

    In this section, we discuss gapless excitations above the ground state.
    Our aim is to obtain an upper bound on the excitation energy $\Delta E(p)$ for momenta $p$ under certain assumptions.
    For Heisenberg models, spin-wave excitations have been widely studied, whereas analogous results for the Hubbard model seem to be limited.
    In the Hubbard model, Momoi~\cite{Momoi}, following the method of~\cite{Stringari,Wagner,HB}, obtained related bounds for the Nambu--Goldstone modes in the magnetic and superconducting phases when the symmetries are spontaneously broken, together with additional assumptions on the dynamical structure factor.
    In the present paper, we work on finite systems and construct a spin-wave excitation using the deviation of the number of fermions from half filling, without assuming either a non-vanishing order parameter or any conditions on structure factors.
    We emphasize that our estimates rely essentially only on commutation relations, and do not require any further assumptions beyond the filling condition stated below.
    It is also worth noting that establishing orthogonality of low energy excitations above infinite-volume ground states calls for careful treatment; see~\cite{Koma1}.

    {The argument is independent of the topics in the previous sections, and we impose lattice translational invariance on the Hamiltonian $H^{(\Lambda)}$. }
    We introduce
    \begin{equation}
        \label{Oetap}
        O_\eta(p):=\sum_{x\in\Lambda}a_x e^{ipx}\quad \mbox{with} \ \ a_x:=\eta_x c_{x,\uparrow}^\dagger c_{x,\downarrow}^\dagger,
    \end{equation}
    where $p$ is a wavenumber vector.
    {Here $\eta_x$ is the same sublattice sign as defined in (\ref{etax}).}
    In order to evaluate the energy of low-lying excitations above a ground state $\Phi_{\rm GS}$,
    we consider a trial state $O_\eta(p)\Phi_{\rm GS}$. The variational energy is given by
    \begin{equation}
        \label{DeltaE}
        \Delta E(p):=\frac{\langle\Phi_{\rm GS},[O_\eta(p)]^\dagger(H^{(\Lambda)}-E_{\rm GS})O_\eta(p)\Phi_{\rm GS}\rangle}{\langle
            \Phi_{\rm GS},[O_\eta(p)]^\dagger O_\eta(p)\Phi_{\rm GS}\rangle}
    \end{equation}
    with the ground-state energy $E_{\rm GS}$. Here, we can also use $[O_\eta(p)]^\dagger$ instead of $O_\eta(p)$. Then, we have

    \begin{theorem}
        \label{GaplessEx}
        Assume lattice translational invariance for the Hamiltonian $H^{(\Lambda)}$ of the attractive Hubbard model.
        {If the number of fermions, $N$,  in a ground state $\Phi_{\rm GS}$ satisfies $|N-|\Lambda||\ge \delta_0 |\Lambda|$ for some $\delta_0>0$,
            then the following bound is valid:}
        \begin{equation}
            \label{DeltaEbound}
            \Delta E(p)\le C_0|p|,
        \end{equation}
        for the excitation energy $\Delta E(p)$ of (\ref{DeltaE}),
        where $C_0$ is a strictly positive constant which is independent of the volume $|\Lambda|$.
        This implies that there exists a gapless excitation above the ground state $\Phi_{\rm GS}$.
    \end{theorem}

    \begin{proof}{Proof}
        Consider first the numerator on the right-hand side of (\ref{DeltaE}).
        Note that
        \begin{eqnarray}
            \langle \Phi_{\rm GS},O_\eta(p)(H^{(\Lambda)}-E_{\rm GS})[O_\eta(p)]^\dagger \Phi_{\rm GS}\rangle
            &=&\langle [O_\eta(p)]^\dagger \Phi_{\rm GS},(H^{(\Lambda)}-E_{\rm GS})[O_\eta(p)]^\dagger \Phi_{\rm GS}\rangle\nonumber\\
            &=&\langle (H^{(\Lambda)}-E_{\rm GS})[O_\eta(p)]^\dagger \Phi_{\rm GS},[O_\eta(p)]^\dagger \Phi_{\rm GS}\rangle\nonumber\\
            &=&\langle [H^{(\Lambda)},[O_\eta(p)]^\dagger] \Phi_{\rm GS},[O_\eta(p)]^\dagger \Phi_{\rm GS}\rangle\nonumber\\
            &=&-\langle \Phi_{\rm GS},[H^{(\Lambda)},O_\eta(p)] [O_\eta(p)]^\dagger \Phi_{\rm GS}\rangle.
        \end{eqnarray}
        By using this identity, the numerator can be estimated as follows:\;\cite{Koma1}
        \begin{eqnarray}
            \label{doublcommbound}
            & &\langle\Phi_{\rm GS},[O_\eta(p)]^\dagger(H^{(\Lambda)}-E_{\rm GS})O_\eta(p)\Phi_{\rm GS}\rangle\nonumber\\
            &=&\langle\Phi_{\rm GS},[O_\eta(p)]^\dagger[H^{(\Lambda)},O_\eta(p)]\Phi_{\rm GS}\rangle\nonumber\\
            &\le&\langle\Phi_{\rm GS},[O_\eta(p)]^\dagger[H^{(\Lambda)},O_\eta(p)]\Phi_{\rm GS}\rangle
            +\langle \Phi_{\rm GS},O_\eta(p)(H^{(\Lambda)}-E_{\rm GS})[O_\eta(p)]^\dagger \Phi_{\rm GS}\rangle\nonumber\\
            &=&\langle\Phi_{\rm GS},[O_\eta(p)]^\dagger[H^{(\Lambda)},O_\eta(p)]\Phi_{\rm GS}\rangle
            -\langle \Phi_{\rm GS},[H^{(\Lambda)},O_\eta(p)] [O_\eta(p)]^\dagger \Phi_{\rm GS}\rangle\nonumber\\
            &=&\langle\Phi_{\rm GS},\bigl[[O_\eta(p)]^\dagger,[H^{(\Lambda)},O_\eta(p)]\bigr]\Phi_{\rm GS}\rangle.
        \end{eqnarray}
        Further, in order to estimate the right-hand side, we introduce unit cells $\mathcal{U}_{x_0}$ so that
        the lattice $\Lambda$ can be written as a sum of the unit cells:
        \begin{equation}
            \Lambda=\sum_{x_0\in \Lambda_{\rm sub}}\mathcal{U}_{x_0},
        \end{equation}
        where $\Lambda_{\rm sub}$ is the sublattice of $\Lambda$, i.e., $\Lambda_{\rm sub}\subset\Lambda$.
        By using this decomposition, the Hamiltonian $H^{(\Lambda)}$ can be also decomposed into a sum,
        \begin{equation}
            H^{(\Lambda)}=\sum_{x_0\in \Lambda_{\rm sub}}h(\mathcal{U}_{x_0}),
        \end{equation}
        where $h(\mathcal{U}_{x_0})$ is the local Hamiltonian such that the support of $h(\mathcal{U}_{x_0})$ contains
        the unit cell $\mathcal{U}_{x_0}$, and the size of the support is slightly larger than the size of the unit cell.

        By substituting the decomposition of the Hamiltonian $H^{(\Lambda)}$ into the double commutator in (\ref{doublcommbound}),
        we have
        \begin{eqnarray}
            \label{DoubleCommuHO}
            \bigl[[O_\eta(p)]^\dagger,[H^{(\Lambda)},O_\eta(p)]\bigr]
            &=&\sum_{x_0\in \Lambda_{\rm sub}}\bigl[[O_\eta(p)]^\dagger,[h(\mathcal{U}_{x_0}),O_\eta(p)]\bigr]  \nonumber\\
            &=&\sum_{x_0\in \Lambda_{\rm sub}}\sum_{x\in \mathcal{V}_{x_0}}\sum_{y\in\mathcal{V}_{x_0}}
            \bigl[a_x^\dagger e^{-ipx},[h(\mathcal{U}_{x_0}),a_y e^{ipy}]\bigr]\nonumber\\
            &=& \sum_{x_0\in \Lambda_{\rm sub}}\sum_{\delta x_0}\sum_{\delta y_0}
            \bigl[a_x^\dagger e^{-ip\delta x_0},[h(\mathcal{U}_{x_0}),a_y e^{ip\delta y_0}]\bigr]
        \end{eqnarray}
        where we have written $\mathcal{V}_{x_0}:={\rm supp}\; h(\mathcal{U}_{x_0})$
        and $x=x_0+\delta x_0$ for $x\in \mathcal{V}_{x_0}$ and $y=x_0+\delta y_0$ for $y\in \mathcal{V}_{x_0}$.
        Since $a_y$ is the generator of the local SU(2) rotation, we have
        \begin{equation}
            \sum_{\delta y_0}[h(\mathcal{U}_{x_0}),a_y ]=0
        \end{equation}
        with $y=x_0+\delta y_0$. Therefore, we obtain
        \begin{equation}
            \bigl[[O_\eta(p)]^\dagger,[H^{(\Lambda)},O_\eta(p)]\bigr]
            =\sum_{x_0\in \Lambda_{\rm sub}}\sum_{\delta x_0}\sum_{\delta y_0}
            \bigl[a_x^\dagger e^{-ip\delta x_0},[h(\mathcal{U}_{x_0}),a_y (e^{ip\delta y_0}-1)]\bigr].
        \end{equation}
        This yields
        \begin{equation}
            \label{DCpbound}
            \langle \Phi_{\rm GS},\bigl[[O_\eta(p)]^\dagger,[H^{(\Lambda)},O_\eta(p)]\bigr]\Phi_{\rm GS}\rangle\le
            C_1 |\Lambda||p|
        \end{equation}
        with a positive constant $C_1>0$ which is independent of the size of the lattice $\Lambda$.

        Next consider the denominator on the right-hand side of (\ref{DeltaE}).
        By using the anti-commutation relations of $c_{x,\sigma}$, one has
        \begin{equation}
            [O_\eta(p)]^\dagger O_\eta(p)-O_\eta(p)[O_\eta(p)]^\dagger=\sum_{x\in\Lambda}(1-n_x).
        \end{equation}
        Therefore, one has
        \begin{equation}
            \langle \Phi_{\rm GS},[O_\eta(p)]^\dagger O_\eta(p)\Phi_{\rm GS}\rangle =
            \langle \Phi_{\rm GS},O_\eta(p)[O_\eta(p)]^\dagger\Phi_{\rm GS}\rangle +|\Lambda| -N \ge |\Lambda| -N.
        \end{equation}
        Since we can choose the ground state $\Phi_{\rm GS}$ such that $|\Lambda| -N$ is positive and of order $|\Lambda|$,
        by combining this with the above bound (\ref{DCpbound}), we obtain the desired bound (\ref{DeltaEbound})
        for the excitation energy $\Delta E$ of (\ref{DeltaE}).

        In the case of $|\Lambda| -N<0$, we use the operator $[O_\eta(p)]^\dagger$ instead of the operator $O_\eta(p)$ of (\ref{Oetap}).
        Then we obtain a similar bound.
    \end{proof}

    Next, we show that a certain spatial symmetry refines the upper bound (\ref{DeltaEbound}) in Theorem~\ref{GaplessEx}
    by the order of $|p|^2$ with a small $|p|$. Let us consider a spatial inversion $\vartheta_{x_0}$
    with the center site $x_0\in\Lambda$. We require that the local Hamiltonian $h(\mathcal{U}_{x_0})$ is invariant under
    the inversion $\vartheta_{x_0}$, i.e., $\vartheta_{x_0}(h(\mathcal{U}_{x_0}))=h(\mathcal{U}_{x_0})$, and that
    the Hamiltonian $H^{(\Lambda)}$ is also invariant. More precisely, we require the following:
    Write $x=x_0+\delta x_0\in\mathcal{V}_{x_0}$ with $\delta x_0\ne 0$.
    Then, there exists a unique site $y=x_0-\delta x_0\in\mathcal{V}_{x_0}$ such that $\vartheta_{x_0}(\delta x_0)=-\delta x_0$.
    Clearly, we can choose the ground state $\Phi_{\rm GS}$ to be invariant under the transformation $\vartheta_{x_0}$.
    Under this assumption, we have

    \begin{theorem}
        \label{GaplessEx2}
        Assume lattice translational invariance and the above spatial inversion symmetry
        for the Hamiltonian $H^{(\Lambda)}$ of the attractive Hubbard model.
        {If the number of fermions, $N$,  in a ground state $\Phi_{\rm GS}$ satisfies that $|N-|\Lambda||\ge \delta_0 |\Lambda|$ for some $\delta_0>0$},
        then the following bound is valid:
        \begin{equation}
            \label{DeltaEbound2}
            \Delta E(p)\le \tilde{C}_0|p|^2,
        \end{equation}
        for the excitation energy $\Delta E(p)$ of (\ref{DeltaE}),
        where $\tilde{C}_0$ is a strictly positive constant which is independent of the volume $|\Lambda|$.
    \end{theorem}

    \begin{proof}{Proof}
        In order to prove this statement, consider the ground-state expectation,
        \begin{equation}
            \mathcal{S}(x_0):=\sum_{\delta x_0}\sum_{\delta y_0}\langle\Phi_{\rm GS},
            \bigl[a_x^\dagger e^{-ip\delta x_0},[h(\mathcal{U}_{x_0}),a_y e^{ip\delta y_0}]\bigr]\Phi_{\rm GS}\rangle,
        \end{equation}
        for the quantity on the right-hand side of (\ref{DoubleCommuHO}). Clearly, this can be decomposed into two parts,
        $\mathcal{S}(x_0)=\mathcal{S}_1(x_0)-i\mathcal{S}_2(x_0)$, with
        \begin{equation}
            \mathcal{S}_1(x_0):=\sum_{\delta x_0}\sum_{\delta y_0}\langle\Phi_{\rm GS},
            \bigl[a_x^\dagger \cos(p\delta x_0),[h(\mathcal{U}_{x_0}),a_y e^{ip\delta y_0}]\bigr]\Phi_{\rm GS}\rangle
        \end{equation}
        and
        \begin{equation}
            \mathcal{S}_2(x_0):=\sum_{\delta x_0}\sum_{\delta y_0}\langle\Phi_{\rm GS},
            \bigl[a_x^\dagger \sin(p\delta x_0),[h(\mathcal{U}_{x_0}),a_y e^{ip\delta y_0}]\bigr]\Phi_{\rm GS}\rangle.
        \end{equation}
        For the second part, in the same way as in the proof of Theorem~\ref{GaplessEx}, we obtain
        \begin{equation}
            \label{calSsbound}
            |\mathcal{S}_2(x_0)|\le C_2|p|^2
        \end{equation}
        with a positive constant $C_2$. Thus, it is enough to estimate $\mathcal{S}_1(x_0)$.
        One notices that $\sum_{\delta x_0}a_x^\dagger \cos(p\delta x_0)$ is invariant under the inversion $\vartheta_{x_0}$.
        On the other hand, the SU(2) invariance implies
        \begin{equation}
            \sum_{\delta y_0}[h(\mathcal{U}_{x_0}),a_y e^{ip\delta y_0}]
            =\sum_{\delta y_0\ne 0}[h(\mathcal{U}_{x_0}),a_{x_0+\delta y_0}(e^{ip\delta y_0}-1)].
        \end{equation}
        Under the inversion $\vartheta_{x_0}$, the quantity of the right-hand side is transformed to
        \begin{equation}
            \sum_{\delta y_0\ne 0}[h(\mathcal{U}_{x_0}),a_{x_0-\delta y_0}(e^{ip\delta y_0}-1)]=
            \sum_{\delta y_0\ne 0}[h(\mathcal{U}_{x_0}),a_{x_0+\delta y_0}(e^{-ip\delta y_0}-1)].
        \end{equation}
        The sum of these two quantities is given by
        \begin{equation}
            2\sum_{\delta y_0\ne 0}[h(\mathcal{U}_{x_0}),a_{x_0+\delta y_0}[\cos(p\delta y_0)-1]].
        \end{equation}
        Putting all together, we obtain
        \begin{equation}
            |\mathcal{S}_1(x_0)|\le C_3|p|^2
        \end{equation}
        with a positive constant $C_3$ which is independent of the volume $|\Lambda|$. By combining this with the bound (\ref{calSsbound})
        for $\mathcal{S}_2(x_0)$, we can obtain the desired bound (\ref{DeltaEbound2}).
    \end{proof}

    \section*{Acknowledgement}
    We would like to thank Hosho Katsura for helpful discussions and comments.
    We also would like to thank S.-Q. Shen for his useful information.
    Y.~G. is supported by JSPS KAKENHI Grant Number 23K12989.
    H.~Y.\ acknowledges support by JST ERATO Grant No. JPMJER2302, Japan and the Special Postdoctoral Researchers Program at RIKEN.


\end{document}